%% file: PETS_2015.tex
\documentclass{acm_proc_article-sp}
\usepackage{amssymb}
\usepackage{epstopdf}
\usepackage[font=small,labelfont=bf,skip=2pt]{caption}
\usepackage{graphicx,subfig}
\usepackage{times,amsmath,amssymb,verbatim}
\usepackage[hyphens]{url}
\usepackage[colorlinks,urlcolor=blue,breaklinks=true]{hyperref}
\usepackage{breakurl}      
\usepackage{cite}     
\usepackage{mdwlist}
\usepackage{enumitem}
\usepackage{tabularx}
\usepackage{paralist}

\graphicspath{{figs/}}

\begin{document}

\title{Defending Tor from Network Adversaries: A Case Study of Network Path Prediction}

\numberofauthors{3} 
\author{               
\alignauthor Joshua Juen\\
      \affaddr{University of Illinois at Urbana-Champaign}\\
      \email{juen1@illinois.edu}  
\alignauthor Aaron Johnson\\
       \affaddr{Naval Research Laboratory}\\
       \email{aaron.m.johnson@nrl.navy.mil}  
\alignauthor Anupam Das\\
       \affaddr{University of Illinois at Urbana-Champaign} \\
       \email{das17@illinois.edu}
\and
\alignauthor Nikita Borisov\\
       \affaddr{University of Illinois at Urbana-Champaign}\\
       \email{nikita@illinois.edu}   
\alignauthor Matthew Caesar\\
       \affaddr{University of Illinois at Urbana-Champaign}\\
       \email{caesar@illinois.edu}    
}

\maketitle

\begin{abstract}

The Tor anonymity network has been shown vulnerable to traffic analysis attacks by autonomous systems and Internet exchanges, which can observe 
different overlay hops belonging to the same circuit. We aim to determine whether network path prediction techniques provide an accurate picture of the threat from such adversaries, and whether they can be used to avoid this threat.
We perform a measurement study by running traceroutes from Tor relays to destinations around the Internet. We use the 
data to evaluate the accuracy of the autonomous systems and Internet exchanges that are predicted to appear on the path using state-of-the-art path 
inference techniques; we also consider the impact that prediction errors have on Tor security, and whether it is possible to produce a useful overestimate that does not miss important threats. Finally, we evaluate the possibility of using these predictions to actively avoid AS and IX adversaries and the challenges this creates for the design of Tor.

\keywords{Autonomous Systems; Internet Exchanges; Tor}
\end{abstract}

\input{intro}
\input{back}
\input{measurement}

\input{measurementvsinf}

\input{results}
\input{related}

\input{conc}

%
{\footnotesize
\bibliographystyle{abbrv}
\bibliography{PETS_2015} 
}
%
%

\end{document}

%% file: intro.tex
\section{Introduction}

The Tor network for anonymous communication~\cite{tor-design} is susceptible to end-to-end timing attacks~\cite{syverson:pet00}, which allow an adversary who observes traffic from a client to the first Tor router and at the same time traffic from the last Tor router to the destination to deanonymize the connection. Both of these paths traverse a number of Internet routers that belong to various organizations, leaving the possibility that a single network operator running an autonomous system (AS) or an Internet exchange  (IX) will be in the position to observe both paths and thus carry out the end-to-end timing attack~\cite{feamster:wpes2004,ccs-EdmanS09,murdoch-pet2007,anonymity-ipx-thesis,usersrouted-ccs13}. This is made more likely by the concentration of Internet traffic at Tier 1 ISPs and high-volume IXes.

To assess the vulnerability of Tor to AS and IX adversaries, it is necessary to predict the paths that traffic takes on the Internet. Previous work characterizing this threat has relied chiefly on AS-level routing predictions~\cite{globecom-QiuG06}. Such predictions are well known to be incomplete and imprecise, producing erroneous path predictions. Our goal is to evaluate the impact of these errors on the anonymity of Tor. In particular, we are concerned with two research questions:
\begin{compactitem}
\item Are AS-level routing predictions suitable for characterizing the threat of AS and IX adversaries in Tor?
\item Can AS-level routing predictions be used to construct Tor paths that avoid AS and IX adversaries (as has been suggested in previous work~\cite{ccs-EdmanS09,lastor})?
\end{compactitem}

To answer these questions, we performed a measurement study, collecting \emph{traceroute} probes from Tor relays to obtain a more accurate picture of Internet paths actually used by network traffic. In comparing results from traceroutes to state-of-the-art path prediction, we found that the prediction accuracy was notably worse than previously measured, despite the fact that we are interested in a simplified prediction problem looking for the \emph{set} of ASes (or IXes) on a path, rather than the exact sequence. The errors include both extraneous ASes and IXes in the prediction that are not seen in traceroutes and, more worryingly, ASes and IXes in the traceroutes that are missing from the prediction. It is possible to produce an overestimate of the AS and IX sets by considering several of the most likely paths produced by the prediction algorithm, rather than just the top one. Such overestimates reduce but do not eliminate the problem of missing ASes and IXes, at the cost of significantly increasing the number of extraneous predictions.

We next analyze the impact of these prediction errors on the vulnerability of Tor to AS- and IX-level adversaries, with the help of a simulator that faithfully reconstructs Tor paths that may have been chosen by a Tor user\footnote{TorPS: \url{http://torps.github.io/}}. We find that AS and IX path prediction significantly overestimates the threat of vulnerability to such adversaries; at the same time, most users do run a significant risk of compromise by an AS-level adversary as determined from the traceroute data, whereas IX-level adversaries affect only a small fraction of paths. 

We then modify our simulator to specifically avoid selecting paths that are vulnerable to AS or IX adversaries based on predictions, as has been previously suggested. We show that this significantly limits the choice of paths and frequently results in \emph{no} paths being available for use while following the Tor practice of maintaining a long-term fixed set of entry guards into the network. This suggests a new consideration for the already complex set of tradeoffs in the design of the mechanisms for selecting and updating the set of entry guards used in Tor~\cite{cogs}; we note that the situation is made worse by the recent move towards using a single entry guard instead of 3~\cite{dingledine2014one}.

On the other hand, we find that many of these failures are a consequence of over-prediction, as we are often able to find suitable non-vulnerable paths in our traceroute data set despite covering only a fraction of the Tor relays. This suggests that a defense based on proactive path measurement, rather than AS path models, is likely to be more practical and offer better security guarantees.

%% file: back.tex
\section{Background}\label{sec:background}
\subsection{Tor}
Tor is a popular system for anonymous communication online. Tor consists of a
network of volunteer \emph{relays} that form an overlay network and forward
traffic sent by users running Tor clients. As of
February 2015, it contains approximately 7\,000 relays and transfers 
around 70\,Gbps of data for user population estimated at over 2\,000\,000.\footnote{\url{https://metrics.torproject.org/}}

Tor uses \emph{onion routing}~\cite{tor-design} to achieve anonymity. A client 
sets up a connection to a destination by choosing a sequence of three relays, conventionally
called \emph{guard}, \emph{middle}, and \emph{exit}, and establishing a \emph{circuit} through the sequence.
The client encrypts a message once for each circuit relay (a process called
\emph{onion encryption}) then sends it through the
circuit, and each relay removes one layer of encryption before forwarding. The final relay sends
unencrypted messages to the destination. The reverse process happens for messages from the
destination to the client. As a result of this process, the client identity is only directly
observable in traffic between the client and the guard relay, and the destination identity is only
directly observable in traffic between the exit relay and the destination. 

In order to be real-time and efficient, Tor does not mix, pad or delay traffic. Therefore it is
vulnerable to attacks based on traffic analysis. For example, an adversary that can observe a
circuit between the client and guard and also between the exit and destination can
correlate the traffic patterns
and deanonymize the connection~\cite{bauer:wpes2007}. 
Thus entities that can observe
parts the underlying network infrastructure, such as Internet Service Providers or Internet
Exchanges, are a serious threat to Tor. Previous work has shown that individual Autonomous Systems
and  Internet Exchanges are in fact frequently in a position to break Tor's
security~\cite{feamster:wpes2004,murdoch-pet2007,ccs-EdmanS09,anonymity-ipx-thesis,usersrouted-ccs13}.
However, almost all of this analysis uses heuristic route-inference techniques whose accuracy may
not be satisfactory. Murdoch and Zieli\'nski~\cite{murdoch-pet2007} do study Tor security against
IXes using
traceroutes from Tor relays, but that analysis is from the UK only and does not consider whether IX adversaries can be avoided during path selection.

\subsection{Internet routing}
Internet routing at the highest level is performed among autonomous systems using the Border
Gateway Protocol (BGP). An AS is a network with an opaque internal routing policy (e.g., using OSPF,
IS-IS, RIP, or iBGP) that routes traffic to and from other networks. BGP is a \emph{path-vector}
routing protocol, that is, neighboring networks advertise the whole AS path that they will use to
send traffic to a given destination. A path is advertised for an IP \emph{prefix} and represents the
path used for all IP addresses sharing that prefix. Path-vector routing enables each AS to make
complex routing decisions based on factors such as individual contracts with other ASes. 

Understanding the behavior of such complex routing policies on the Internet is a challenging
problem. Routers just propagate the routes that they provide for a given neighbor to use, and so
different Internet vantage points reveal
different subsets of global routing behavior. Sources of routing data include the Route Views 
Project\footnote{\url{http://www.routeviews.org/}}, which provides
BGP routing information from many large ASes, and CAIDA Archipelago\footnote{\url{http://www.caida.org/projects/ark/}}, which provides
and analyzes traceroute data from three teams of 17--18 monitors distributed worldwide.
Gao describes how to use such data to infer Internet routes~\cite{Gao2001}. Gao's method uses
heuristics to classify the observed connections between ASes by their economic relationship
(viz.\ customer-to-provider, provider-to-customer, peer-to-peer, or sibling), and
then shortest-path valley-free routing is used to infer the route between two hosts. Qiu and Gao
improve the accuracy of this technique by incorporating the observed advertised BGP
paths~\cite{globecom-QiuG06}. In addition, they describe how to infer a set of possible paths rather
than just one. Their results show that these techniques can infer the exact correct AS path for
60\% of evaluation ASes; furthermore, the exact path is found within the top 5 predicted possible paths
for 83\% ASes and within the top 14 paths for 86\% ASes.

Many links between ASes occur at Internet Exchanges. These are facilities that provide
space and infrastructure for ASes to locate routers and establish connections.
Ager et al.~\cite{ager2012anatomy} describe how the largest IXes may provide links among hundreds
of ASes and carry petabytes of traffic per day. Augustin et al.~\cite{AugustinIXP} describe
how IXes on Internet routes can be detected using traceroutes and an index of known IXes and their
IP prefixes. They identify 44\,000 peering relationships between ASes at IXes.
Each peering
between two ASes indicates that \emph{some} traceroute passed directly from one AS to another
through an IX. Discovering such links can improve the accuracy of AS path inference techniques. However, as we will observe, it doesn't discern among different router-level paths taken between the
same two ASes, which may pass through different IXes.

The traceroute tool is extraordinarily useful in measuring routing behavior on
the Internet.
There are many variations of the basic algorithm~\cite{imc-Luckie2008} which provide
different levels of success depending on the traffic engineering (e.g. filtering and load balancing)
that occurs en route. In addition to such problems with traceroute itself, it is not always
straightforward to make inferences about Internet paths from a traceroute. For example,
Mao et al.~\cite{sigcomm-mao2003} describe the difficulties of inferring an AS-level path from
traceroutes, which include that different iterations of a single traceroute might take different
paths, that reported IP addresses may be from a network interface other than the one that actually
received it, and that mapping from IP address to AS number is non-trivial due to inaccurate WHOIS
information. Augustin et al.~\cite{AugustinIXP} discuss similar issues in inferring the presence
of IXes from traceroutes. Nevertheless, traceroutes do provide a generally accurate picture of how 
packets are actually routed and serve as an important comparison point to AS-level path predictions.

%% file: measurement.tex
\section{Mapping Network Adversaries}

\subsection{Measuring Internet Paths}{\label{measurement}}
 
\subsubsection{Generating Traceroutes}

Our measurement study consists of running traceroutes from Tor relays to various destinations in the Internet. We use the
\emph{scamper}\footnote{\url{http://www.caida.org/tools/measurement/scamper/}} network tool, which probes multiple destinations in parallel, and uses techniques to accurately discover the Internet path traversed by packets in the presence of multi-path load balancing~\cite{scamper, augustin2006avoiding}. 

For our measurements, we extracted the set advertised destination IP prefixes from the September 2013 RIB dumps from Route Views. Each relay running the measurements picks a random IP address within each of the $\sim$500K prefixes and performs a traceroute to that destination. We also collected traceroutes to Tor relay destinations as well as a scan of all /24 IPv4 subnets, but this data was not used for the analysis in this paper. We focus on advertised prefixes to make analysis more tractable. We expect addresses within a prefix to use the same or similar routes, and our analysis of CAIDA's traceroutes to all /24 IPv4 subnets~\cite{caidaData} found that 81\% of the time traceroutes destined to the same routable prefix traversed the same set of ASes. 
Our measurement scripts are available for public review.\footnote{URL removed for blinded review}

\subsubsection{Processing Traceroutes}
                                                         
We next process the traceroutes to determine which ASes and ISPs an Internet path has traversed. First, we filter out traceroutes that do not successfully reach the destination. Note that because we use randomized destinations, in many cases the destination may not exist or may be down; indeed, only a small fraction (8\%) of probes reaches their target. However, 49\% reach the AS of the destination, as determined by the MaxMind GeoIP database~\cite{geoip}.
                     
We further find that 94\% of the traceroutes are missing some hops from the path. In some cases, we believe this is caused by routers close to the probe source rate limiting their ICMP responses. To address this, we perform \emph{route stitching}, where gaps in a traceroutes are filled by path segments observed in other traceroutes. For example, if we see a path ``A B C D E'' and another path ``A B * D F,'' where ``*'' denotes a missing hop, we can repair the second path by inferring that the third hop must have also been C in this case. To minimize inaccuracies introduced by this repair mechanism, we only consider path segments that originate from the same host, and which are contained within the same batch of 64K traceroutes, which typically occur within an hour or two of each other. We validated this approach on complete paths and found that stitching would have given us the correct AS path result 96\% of the time.

We then compute the ASes corresponding to each IP in the path using the GeoIP database. Similar to Mao et al.~\cite{sigcomm-mao2003}, we consider the corresponding AS path complete if the traceroute reached the AS of the destination and there are no missing hops in the path on the boundary between ASes. For example, an AS path ``AS1 AS1 * AS1 AS2 AS3'' is considered complete, because the missing hop is contained entirely within AS1, whereas ``AS1 AS1 * AS2 AS3'' is considered incomplete. Overall, 28\% of the traceroutes yield a complete AS path. We discard the other traceroutes from our analysis. We also identify an IX as on the path if the path contains an IP address from the list of known IP addresses of IX points as outlined in the following section.

\subsection{Inferring Path ASes and IXes}\label{sec:prediction}

We are interested in comparing the AS and IX adversaries identified from traceroute data compared to AS and IX adversaries inferred from AS maps which are much easier to attain and maintain. We predict AS paths from source to destination using Gao's algorithm~\cite{Gao2001} to classify
relationships and Qiu and Gao's algorithm\cite{globecom-QiuG06} to infer the top $k$ paths (for $k$ = 1 to 5). While advances have been made in classifying AS link relationships~\cite{imc-caida13}, we find that
matching route information broadcasts is more accurate than using graph based methods based solely on AS relationships \cite{anonymity-ipx-thesis}. It is known that AS relationships are difficult to classify especially at the highly interconnected core of the AS graph. Violations in the valley-free principle from advertised routes often indicate erroneous AS relationship classification especially through top tier ASes. Therefore Qiu and Gao's method of prepending advertised routes to complete paths yields accurate results even with incorrectly classified AS relationships at the core of the Internet. Since the prepended hops are almost entirely easily classified customer-to-provider hops at the bottom of the AS graph, improving the AS relationship classification of the top-level ASes does little to improve overall AS path prediction accuracy.

To predict the presence of IXes, we recreate the work of Augustin et al.\cite{AugustinIXP}.  We scraped Packet Clearing House\footnote{\url{http://www.pch.net}} and the Peering Database\footnote{\url{https://www.peeringdb.com}}
 in February of 2014 creating a list of 732 Internet exchange points and their known prefixes. We parsed over 200 million
traceroutes from February and March 2014 collected from both the CAIDA routed IPv4 database~\cite{caidaData} and the iPlane project\footnote{\url{http://iplane.cs.washington.edu/}} to map roughly 130\,000 Internet
exchange point peerings for our subsequent test data. This number is roughly twice the number of links found by Augustin et
al. in 2009, which is unsurprising considering the trend for ASes to peer at IX points. This list of source AS, destination AS, and IX number is used
to identify potential IX points on AS-level paths throughout our experiment. The list of IXes and prefixes is used to positively identify IX points in
the IP traces.

%% file: measurementvsinf.tex
\begin{figure*}[t!]
	\begin{minipage}[c]{\textwidth}
		\begin{minipage}[t]{\textwidth}
			\begin{minipage}[b]{0.49\textwidth}
\includegraphics[width=\columnwidth]{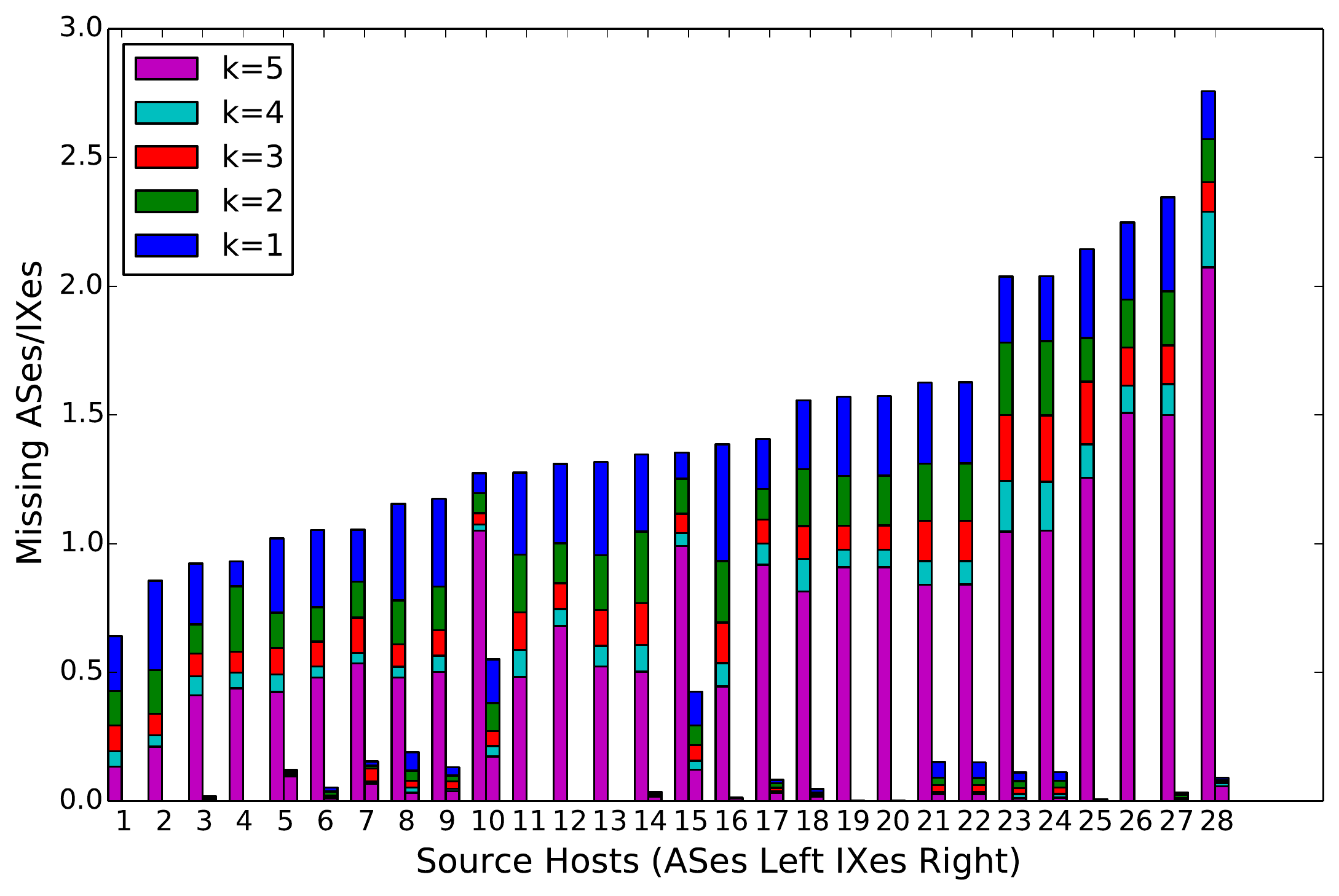}
\caption{Avg Missing ASes and IXes Per Host}
\label{fig:misas}
			\end{minipage}
			\hfill
			\begin{minipage}[b]{0.49\textwidth}
\includegraphics[width=\columnwidth]{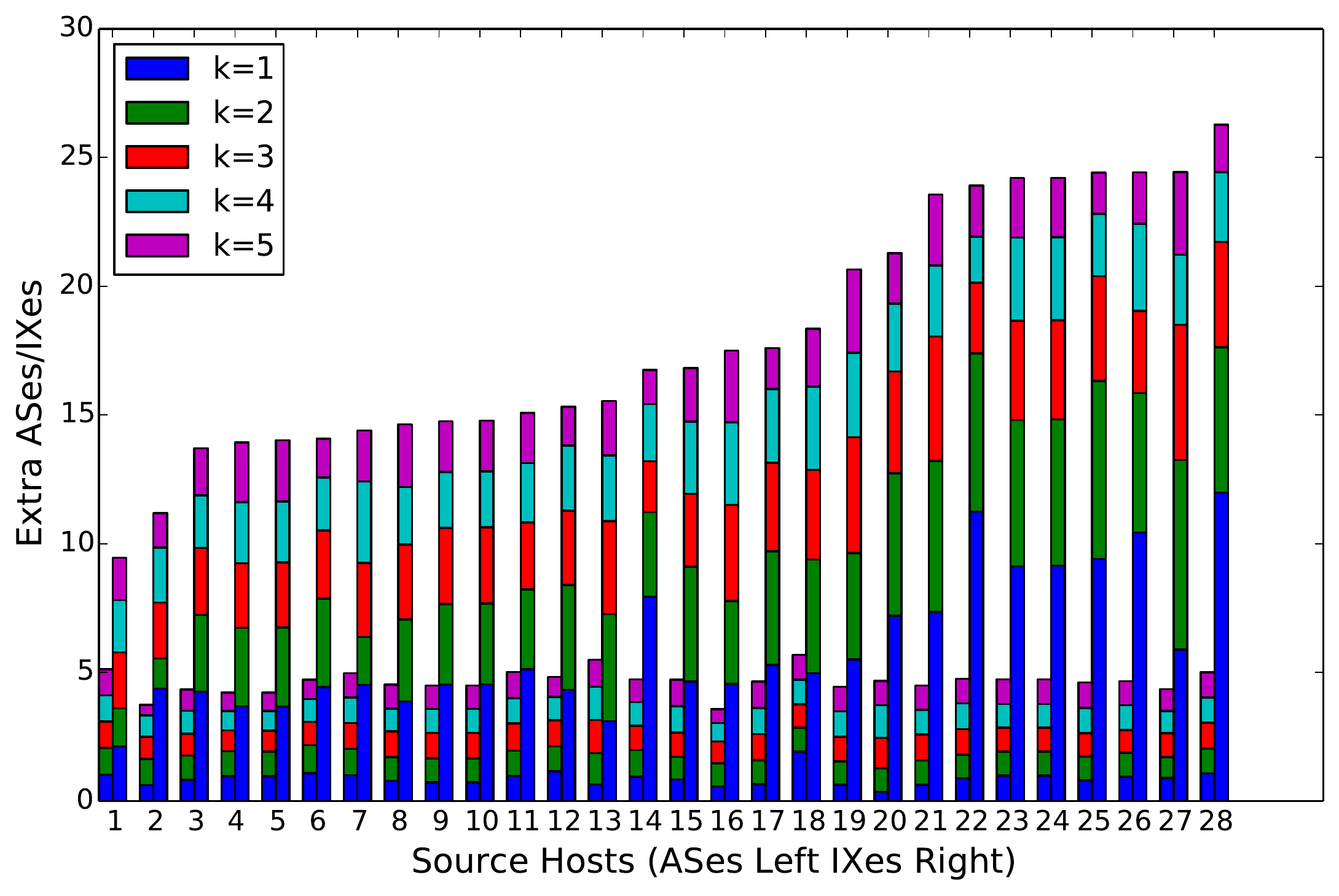}
\caption{Extra ASes and IXes Per Host}
\label{fig:extraas}
			\end{minipage}
		\end{minipage}
		\vspace{3mm}
		\begin{minipage}[t]{\textwidth}
			\begin{minipage}[b]{0.3\textwidth}
\includegraphics[width=\columnwidth]{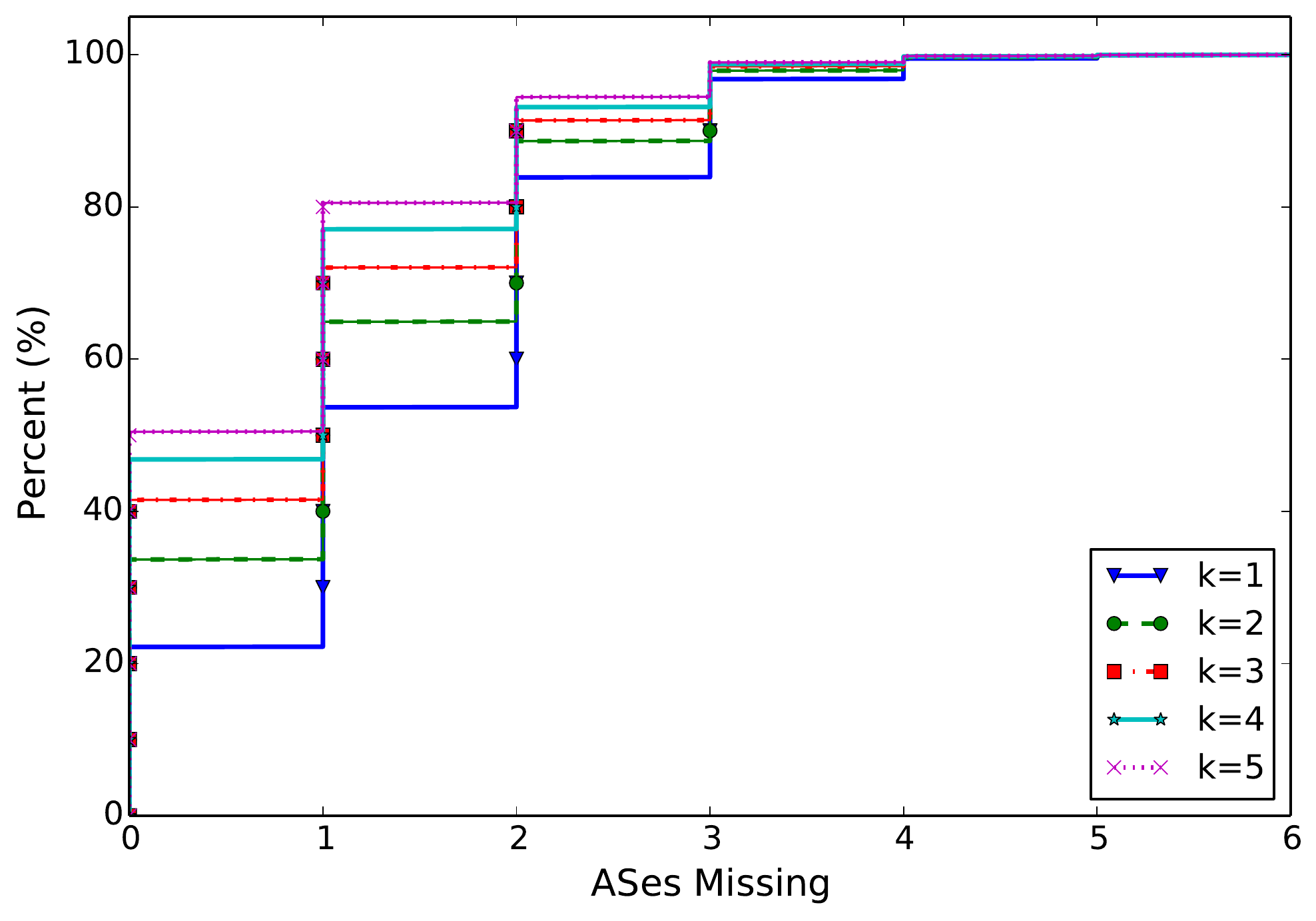}
\caption{CDF of Missing ASes over all Traces}
\label{fig:misascdf}
			\end{minipage}
\hfill
			\begin{minipage}[b]{0.3\textwidth}
\includegraphics[width=\columnwidth]{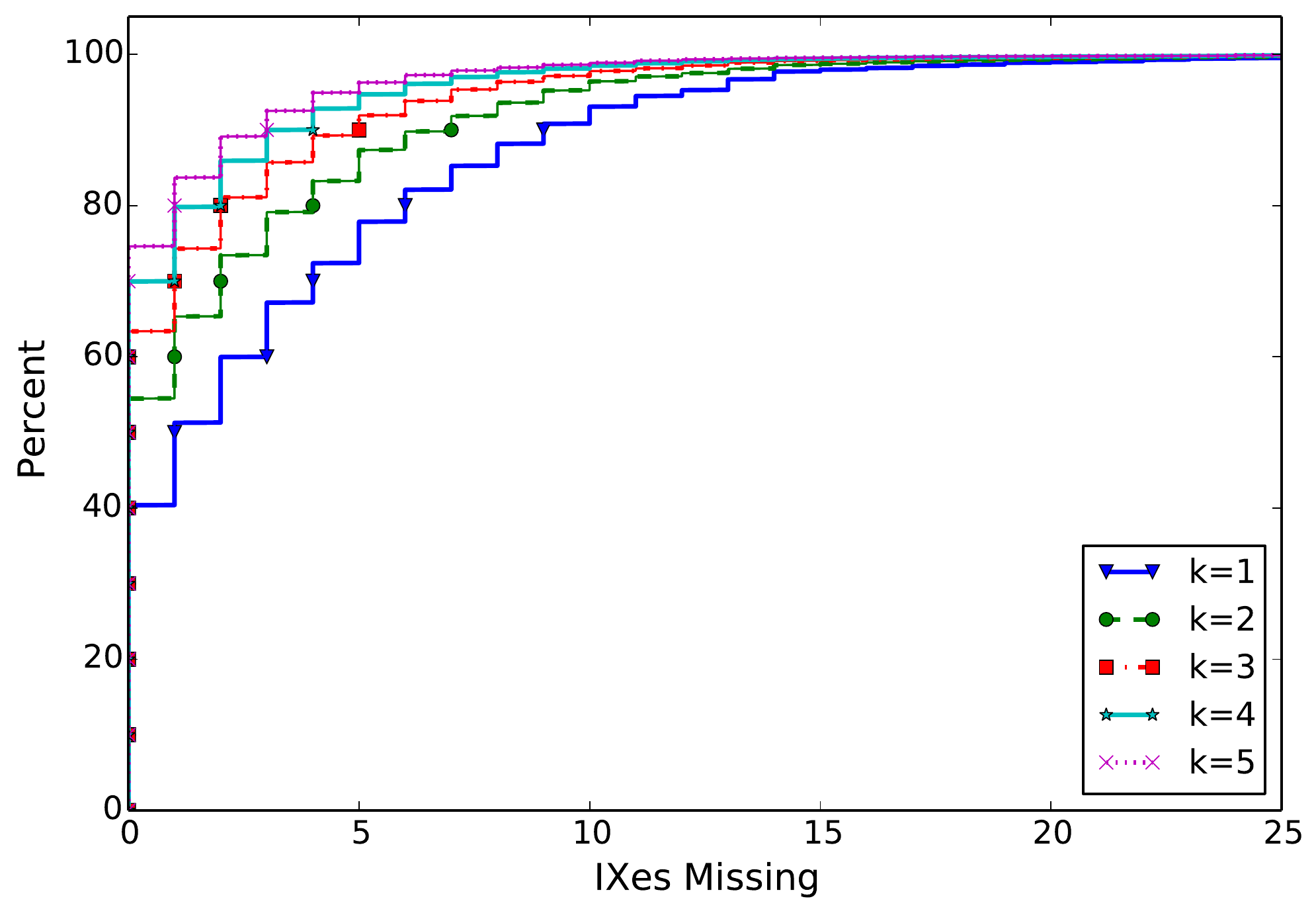}
\caption{ CDF of Missing IXes over all Traces }
\label{fig:ixesmissing}
			\end{minipage}
\hfill
			\begin{minipage}[b]{0.335\textwidth}
\includegraphics[width=\columnwidth]{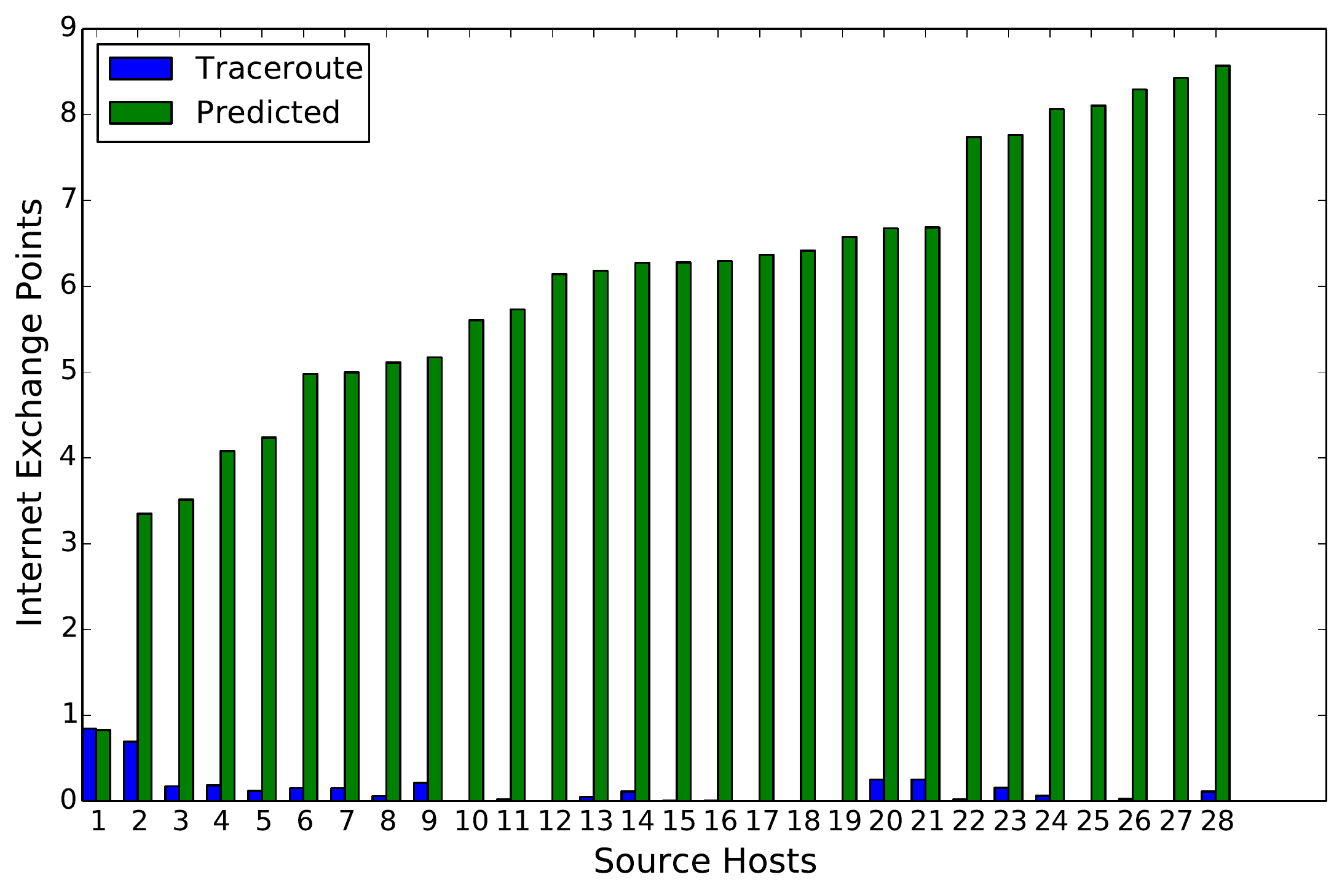}
\caption{Observed Internet Exchange Points}
\label{fig:ixesseen}
			\end{minipage}
		\end{minipage}
	\end{minipage}
\end{figure*}

\section{Measurement versus Inference}

\begin{table}[tb]

\centering
\begin{tabular}{|l|rrr|}
\hline & & & \\ [-1.5ex]
 & Min & 
Max & Total \\
\hline & & &\\ [-1.5ex]
Traces & 392\,334 & 
1\,195\,577 & 17\,233\,153 \\
IP Reached & 27\,254 & 
81\,999 & 1\,183\,427 \\
AS Reached & 204\,836 & 
618\,224 & 8\,890\,142 \\
Repaired & 26\,375 & 
876\,386 & 9\,643\,679 \\
Whole IP Path & 0 & 
16\,694 & 253\,058 \\
Whole AS Path & 11\,285 & 
397\,533 & 5\,350\,713 \\
Probed Hops & 7\,005\,020 & 
20\,624\,809 & 297\,301\,238 \\
Responsive Hops & 1\,594\,449 & 
7\,041\,819 & 109\,046\,956 \\
\% Hops Responded & 19 & 
55 & 37 \\
Inferred Paths & 10\,693 & 
220\,449 & 4\,367\,097 \\
\hline
\end{tabular}
\vspace{1mm}
\caption{Traceroute experiment statistics (28 hosts)}
\label{table:tracestats}
\end{table}

We first investigate how closely ASes and IXes identified through inference correspond to ASes and IXes measured with traceroute. We conduct analysis on 17 million traceroute measurements obtained from 28 Tor relay servers from January 19th to January 26th 2014 as summarized in Table \ref{table:tracestats}\footnote{We extend sincere thanks to the volunteer Tor relay operators who assisted us with this study.}.
Our 28 servers included many of the largest Tor relays and cover a portion of the Tor network which includes 23\% of guard node capacity and 26\% of exit node capacity. Thus, their measurements
can give us good insight into how traffic is routed in and out of Tor.
Of these 17 million traces, only about 1 million reached the target IP address with roughly 250\,000 complete paths with no missing IP hops. We find that roughly 9 million paths can have some hops filled in by using the repair techniques presented in section 3. We map each IP address to AS numbers using the MaxMind GeoLite ASN database taken from January 15th 2014\cite{geoip}. The AS-level paths are then parsed to remove routing loops, duplicate hops, and missing hops directly preceded by and followed by the same AS. After processing, we obtain 5.3 million complete AS-level routes.

\subsection{Identifying AS Adversaries}

We first investigate the accuracy of inferring ASes between an arbitrary source/destination AS compared to the ASes identified in our collected traceroutes. The analysis of path prediction accuracy is conducted on traceroutes collected during January 19--26, 2014 giving 5.3 million traces contain 450\,000 unique AS source and destination pairs. We divide the traceroute data into 24-hour windows. Routing table dumps are downloaded from each server from the Route Views project from the time closest to the 12th hour of the window. Each day window contains an average of 15 prefix table dumps with between four to six gigabytes of route information broadcasts. Using Qiu and Gao's model we predict the top $k$=5 paths for roughly 400\,000 of these pairs with the rest failing due to either the source or destination AS missing from the Route-Views routing tables. The 400\,000 successful path inferences cover 4.5 million of our 5.3 million traces with AS paths. We consider the inference identifying the correct path if it matches the AS path seen in the traceroute for any of the top $k$ paths considered. 

Figure \ref{fig:misascdf} is a CDF showing the number of ASes seen in the traceroute but missed by the top $k$ predicted paths. Zero missing ASes correspond with a correct path prediction for at least one of the $k$ paths. Using only the top path from the prediction, yields roughly 20\% prediction accuracy with a decreasing return for higher levels of $k$ and a maximum accuracy of 48\% when considering the top 5 paths. This accuracy is far lower than the 83\% accuracy for the top 5 paths attained by Qiu and Gao; however, their validation was conducted using Route Views data as the ground truth and not traceroute data~\cite{Gao2001}. We surmise the lower accuracy is due to a combination of known error sources both from the increase in prevalence of IX peering~\cite{ager2012anatomy} and inherent errors in traceroute measurements from actual AS-level paths \cite{zhang2011framework}. We finally note that the overall accuracy of prediction is similar to our extensive analysis of accuracy against CAIDA traceroutes demonstrating that the Tor network would benefit from improvements to path prediction in the general case~\cite{anonymity-ipx-thesis}.

\subsection{Identifying IX Adversaries}

Given an AS path, we can identify the set of potential IXes that could occur on this path by considering which IXes can be used for each AS--AS hop, as discussed in Section~\ref{sec:prediction}. Figure \ref{fig:ixesmissing} compares the set of potential IXes for the top $k$ predicted AS paths to the set to the IXes identified by IP address prefix in traceroutes. Once again, a value of zero indicates that all IXes have been identified in the inference. We find that the top path identifies roughly 40\% of the IXes and the top five paths identify roughly 74\% of the IXes. 

We expect traceroutes to provide a much more accurate picture of which IXes were involved on a path than AS path predictions. A pair of ASes will often have multiple peering points, depending on the geographic location the source and destination; as a result, only a fraction of traffic between the two ASes will use a given IX. In Figure~\ref{fig:ixesseen}, we compare the set of IXes on a traceroute to the predicted set of IXes that could be used at each AS--AS hop in the traceroute. We see that while most traceroutes do not traverse \emph{any} IX, the AS--AS hops result in 3--8 potential IXes on average. This demonstrates the limitations of using only AS-level information to infer IXes on a path.
  
\begin{figure*}[t!]
	\begin{minipage}[c]{\textwidth}
		\begin{minipage}[t]{\textwidth}
			\begin{minipage}[b]{0.49\textwidth}
\includegraphics[width=\columnwidth]{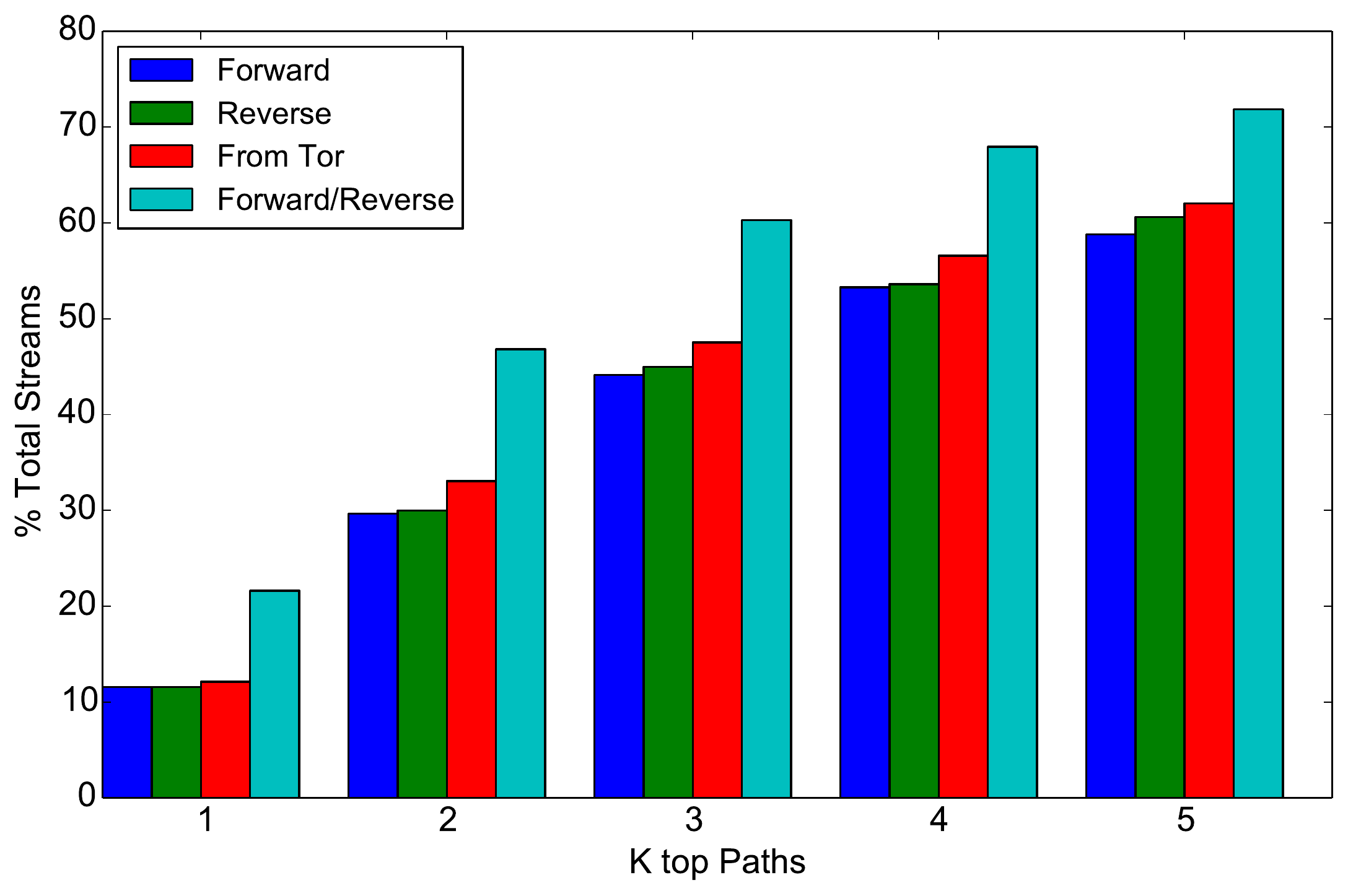}
\caption{Directional AS Compromises for K Top Paths}
\label{fig:dascomp}
			\end{minipage}
			\hfill
			\begin{minipage}[b]{0.49\textwidth}
\includegraphics[width=\columnwidth]{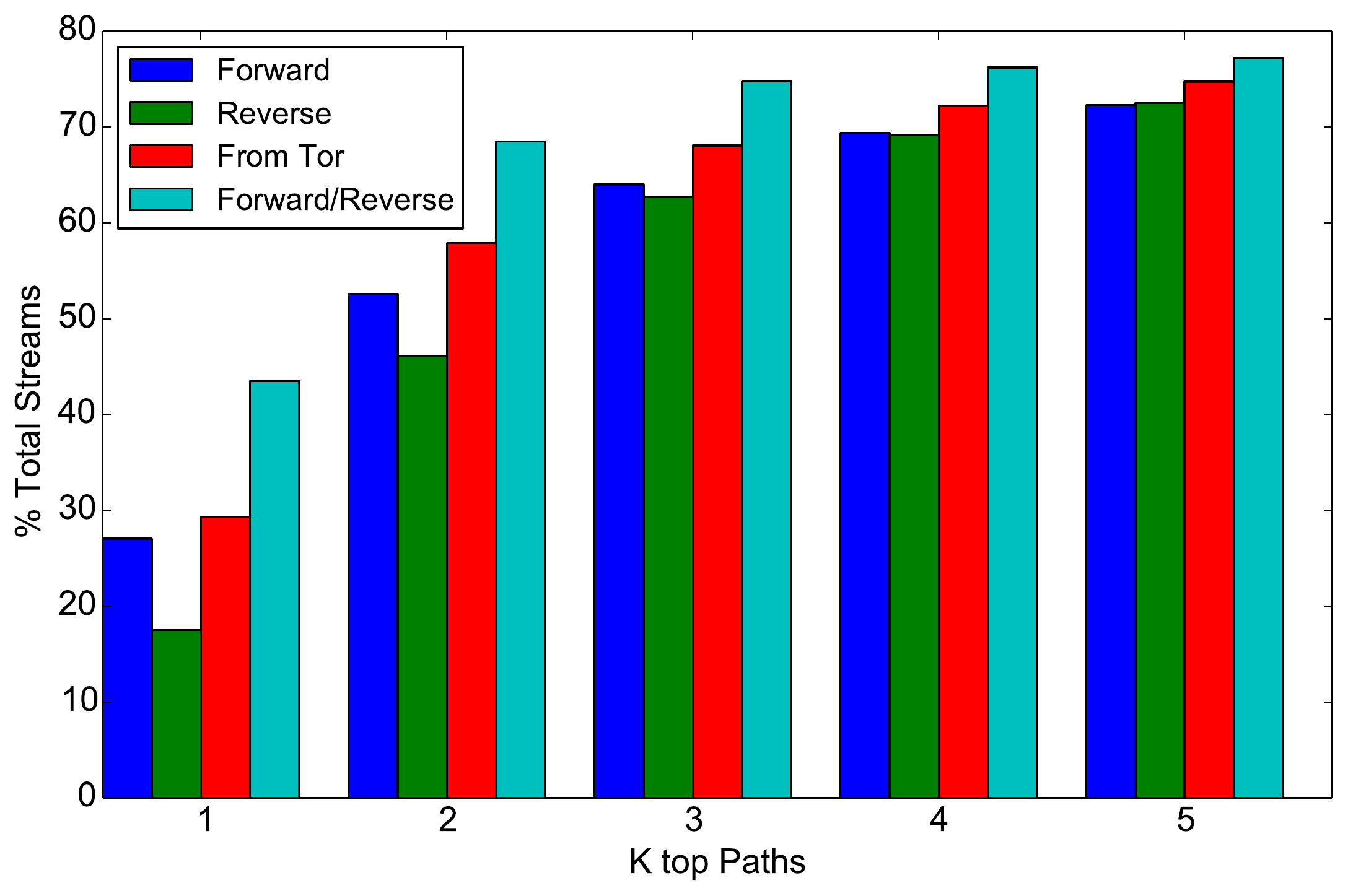}
\caption{Directional IX Compromises for K Top Paths}
\label{fig:dixcomp}
			\end{minipage}
		\end{minipage}
		\vspace{3mm}
		\begin{minipage}[t]{\textwidth}
			\begin{minipage}[b]{0.49\textwidth}
\includegraphics[width=\columnwidth]{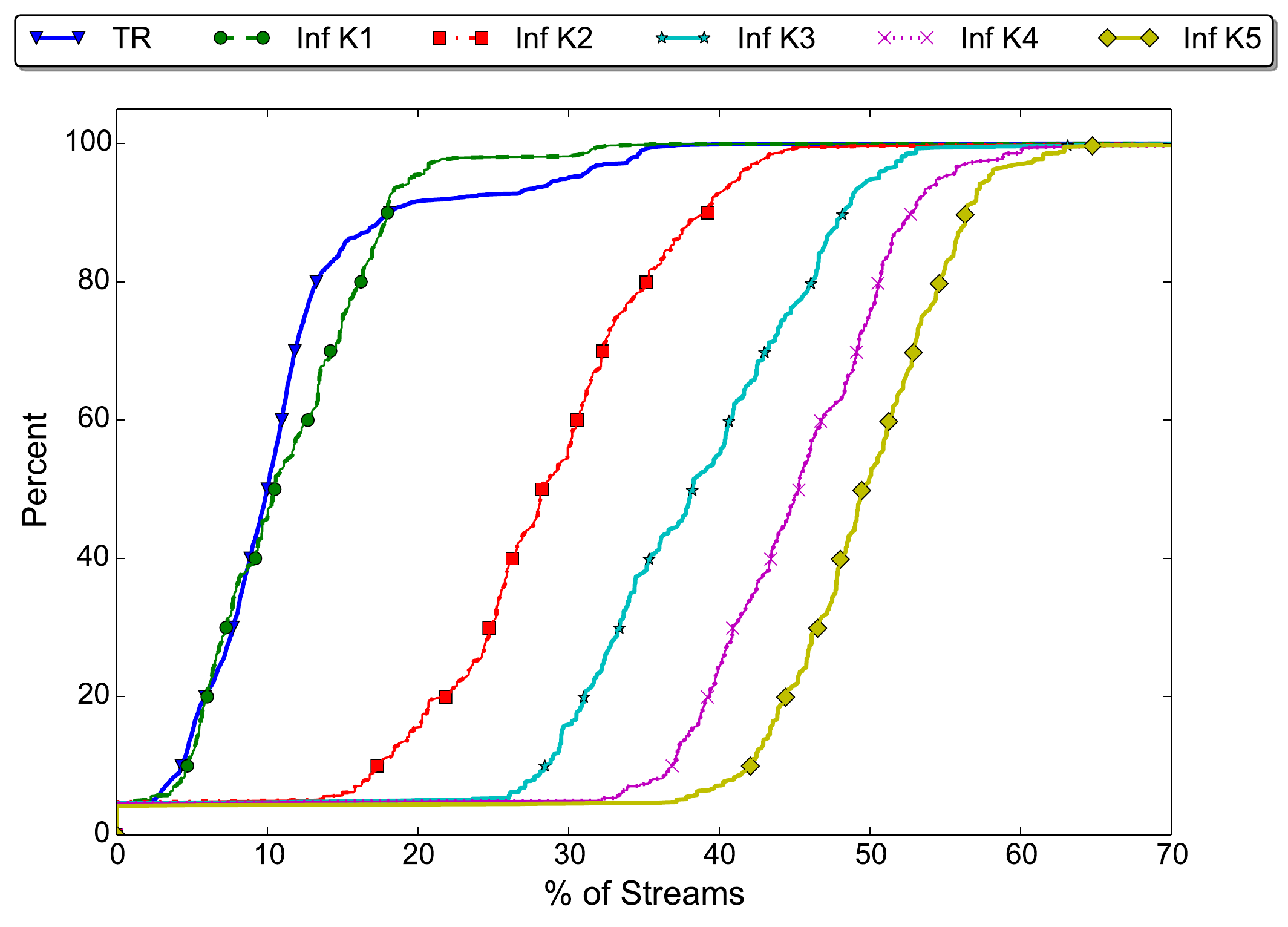}
\caption{AS Compromises Measured and Inferred}
\label{fig:ascomprates}
			\end{minipage}
\hfill
			\begin{minipage}[b]{0.49\textwidth}
\includegraphics[width=\columnwidth]{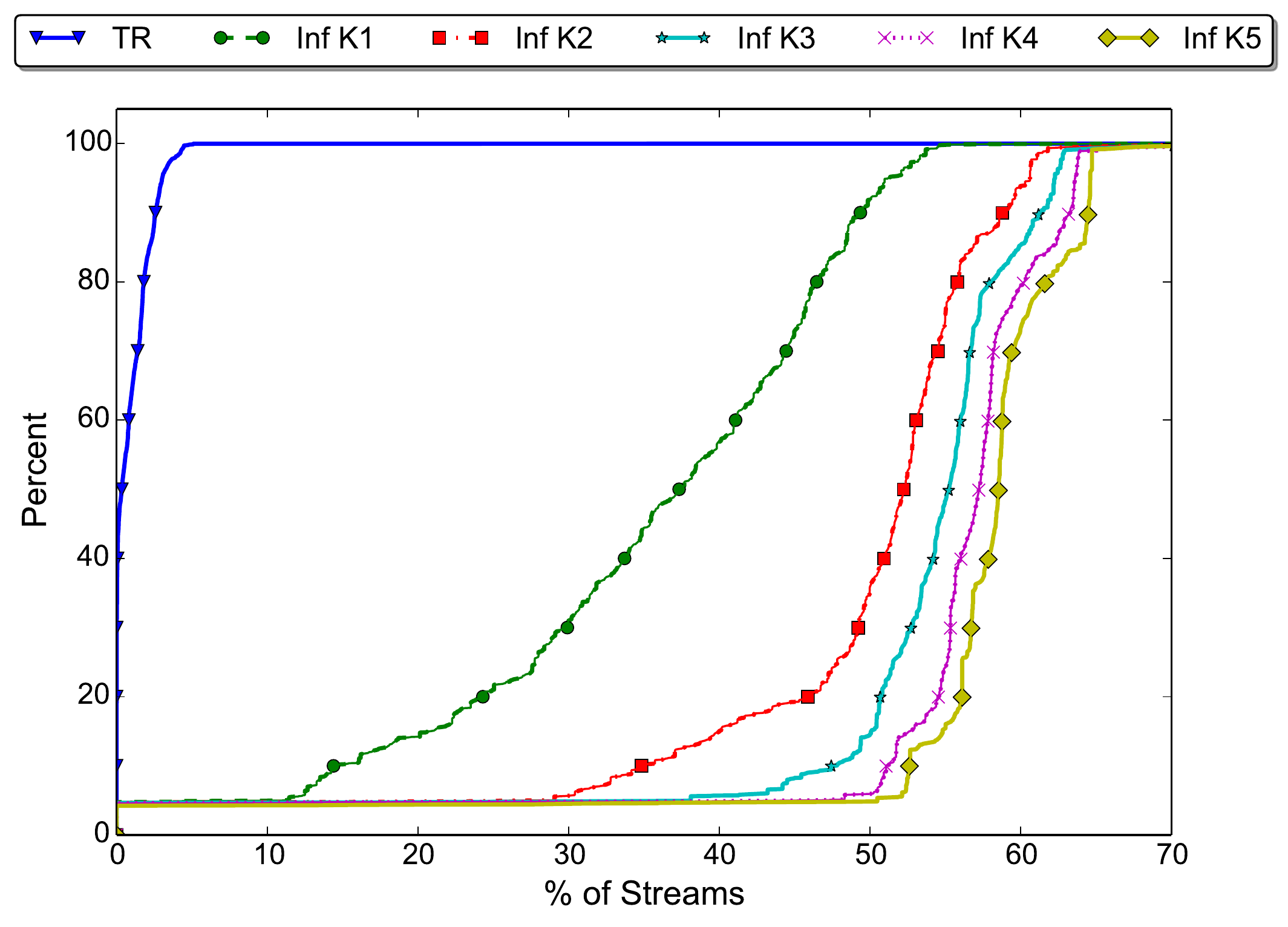}
\caption{IX Compromises Measured and Inferred}
\label{fig:ixcomprates}
			\end{minipage}
		\end{minipage}
	\end{minipage}
\end{figure*}

\subsection{Choosing $k$ Top Paths}

Choosing the $k$ top paths to consider when predicting AS and IX points presents an important tradeoff between missing ASes/IXes versus severely overstating the number of ASes/IXes on a given path. Figures \ref{fig:misas} and \ref{fig:extraas} show the missing AS/IXes and extra AS/IXes seen by the path prediction algorithms for the $k$ top paths from $k$ = 1 to 5. For ASes, we see diminishing returns for missing ASes for larger values of $k$ with false positives increasing quickly for larger values of $k$. Overall, the average attainable missing AS accuracy is close to 1 for the top path decreasing down to .6 for the top 5 paths.
 
We compare the IXes found using vulnerable AS--AS hops from the inferred AS paths directly to the IXes identified by prefix in the traceroutes. In general, very few IX points were seen in the traceroutes. In most hosts, IX identification is helped very little by increasing the top paths with the average missing IX of about .2 per hop. This is unsurprising because if there are no IX points in the traceroutes, then there can be no missing IXes in the inference. Unfortunately, the false positives for IX points are problematic with linearly increasing averages ranging from 10-25 for each of our hosts illustrating the need for better methods in identifying IX points.

For AS adversaries, a $k$ value of 1 or 2 seems most appropriate to identify most AS adversaries without causing too many false positives. Higher values of $k$ give lower rates of return while causing a linear increase in false positives. Identifying IX adversaries is much more problematic. Since the traceroutes identify very few IX adversaries to begin with, a $k$ value of 1 appears to work well. The inaccuracy of the method can be seen in the false positives which also increase linearly with $k$ but greatly over-predict the number of adversaries even with a $k$ value of 1. The inaccuracy of AS and highly inaccurate IX prediction could potentially cause serious problems when designing a system of AS/IX independence in Tor. We analyze the effects of this inaccuracy in the following sections.

%% file: results.tex
\section{AS and IX Adversaries in Tor}

Errors in path prediction call into question previous work that has used path prediction to both
evaluate the security of Tor and propose changes to Tor's path selection based on path
predictions. Understanding the effect of the errors uncovered by our traceroute measurements
requires taking into account the specific properties of Tor.

We accomplish such an analysis by simulating the Tor protocol and network at a high level. We use
and adapt the Tor
Path Simulator (TorPS)\footnote{\url{https://github.com/torps}} to perform Monte Carlo simulation
of Tor path selection by a single client. By using the hourly network ``consensuses'' and server
``descriptors'' archived by CollecTor\footnote{\url{https://collector.torproject.org/}}, we can
recreate the state
of the Tor network over the period we run our simulations, including features such as the number,
bandwidths, and addresses of Tor relays available in any given hour. We simulate ``typical'' user
activity using the recorded volunteer trace of Johnson et al.~\cite{usersrouted-ccs13}, which
includes user behaviors such as web search and webmail on a plausible daily schedule.
Over the course of a week, this schedule results
in 2632 \emph{streams} (i.e., TCP connections over Tor), each to one of 205 distinct IP addresses
occupying 168 unique ASes, on either port 80 or 443.
Finally, we run simulations using the most common client ASes as measured by Juen in Fall 2011
\cite{anonymity-ipx-thesis}.

Simulating path selection in Tor allows us to estimate which Internet hosts a user's
traffic is likely to flow over in a typical use case. Then we can use our traceroute data
to determine the specific Internet routes that traffic would take and evaluate the resulting
security. Specifically, we provide new estimates for how often a Tor stream flows through the same
AS or IX
between the client and the guard and between the destination and the exit.
When this happens, the AS or IX is in a position to deanonymize the client. This issue
was previously studied only using inferred AS
paths and IX sets.

In addition, using this method we provide an
improved evaluation of the repeatedly-proposed~\cite{feamster:wpes2004,ccs-EdmanS09} modification to
Tor to use AS/IX path inference to choose relays that are \emph{path independent}, that is, that
result in paths for which the same AS or IX cannot observe both the client and the destination.
We modify TorPS to produce the first simulator for path-independent Tor (to our knowledge) that
reproduces how path selection occurs over time, including features that have the potential to
significantly alter the effectiveness of the path-independence requirement, such as guard lists and
circuit reuse. We apply our traceroute measurements to the results of these simulations to evaluate
the effectiveness of path inference as a basis for path independence in Tor.

\subsection{Vanilla Tor} \label{subsec:vanilla}

All of our Tor simulations run over the week of January 19--25, 2014. When producing and analyzing
these simulations, we generally use the same data sources and inference
algorithms as in Sec.~\ref{sec:prediction} to produce AS path inferences, AS-level IX inferences,
and traceroute IX inferences. We use daily AS-path inferences conducted from January 19th-25th 2014 compared to the traces from each day of the simulation week. We also use the daily Route Views prefix-to-AS datasets to determine routed prefixes and to map IPs to ASes. When analyzing our simulations using traceroutes, we use all of the traceroute measurements gathered during the week of January 19th-25th 2014. In our analysis we match a traceroute to a pair of communicating hosts in Tor if the source prefix and destination prefix match. 

We first conduct a simulation using the default Tor path selection algorithm. We consider clients coming from 50 of the top 200 most common client ASes (as measured by Juen~\cite{anonymity-ipx-thesis}). Each AS advertises hundreds of possible prefixes in the Route Views data. We select at random twenty prefixes per client AS for a total of 1\,000 client prefixes for the simulations. The simulator runs 10\,000 repetitions of simulated traffic using input data from the week of January 19th--25th 2014 yielding over 24 million traffic streams per client prefix with 18.2 million unique streams. We identify the presence of AS and IX adversaries using AS-path inference with the top k paths (k=1 to 5) and our collected traceroute data from January 19th--January 25th. In total, we have inferred path information for an average of 18 million streams per client prefix (18 billion total) and traceroute information for an average of 112\,000 streams per client prefix (14 million total).

\subsubsection{Inferred Adversaries}

We first look at the percentage of simulated Tor paths which have the same AS or IX on both the client-to-guard path and the exit-to-destination path using only the inferred paths. We look at the percentage of compromised paths considering the set of ASes and IXes in the forward direction (client to destination), the reverse direction, the forward and reverse directions combined. We also consider the direction of streams leaving Tor; i.e., from the guard to the client and from the exit to the destination. This matches the direction of our traceroute measurements from Tor relays to external IP prefixes and allows us to compare the predicted paths with traceroute data, without errors being introduced due to asymmetric Internet paths that traverse a different set of ASes and IXes. We call this the Tor path.

Figures \ref{fig:dascomp} and \ref{fig:dixcomp} show the percentage of inferred ASes and IXes for each direction and top k paths averaged over all 18 billion inferred streams. Considering only the top path, we see 11.6\%, 11.6\%, 12.1\% and 21.6\% AS compromise rates for the forward, reverse, Tor, and forward/reverse paths respectively. We see a significant increase in AS adversaries when considering more paths topping at 58.8\%, 60.6\%, 62.0\%, and 71.8\% when considering the top 5 paths for the forward, reverse, Tor and forward/reverse paths respectively. We notice little difference between the compromise rates of the Tor paths versus the forward or reverse. As expected, the forward and reverse combined represents a higher inferred compromise rate since we consider two sets of ASes per path. The forward/reverse has roughly a 10\% greater rate of compromise for the top path and roughly a 20\% greater compromise rate when k is varied from 2 to 5. For the top path, the IX compromise rates were higher with 27.0\%, 17.5\%, 29.3\% and 43.5\% for the forward, reverse, Tor and forward/reverse paths respectively. These increased rapidly at first leveling off to 72.3\%, 72.5\%, 74.7\% and 77.2\% for the top 5 paths. Once again, the forward/reverse paths contain more potential IX adversaries due to considering more paths. There is little significant difference between the compromise rates of the forward paths and the Tor paths. We also note that the number of inferred potential adversaries greatly increases when considering a higher number of top k paths. 

\begin{figure*}[t!]
	\begin{minipage}[c]{\textwidth}
		\begin{minipage}[t]{\textwidth}
			\begin{minipage}[b]{0.32\textwidth}
\includegraphics[width=\columnwidth]{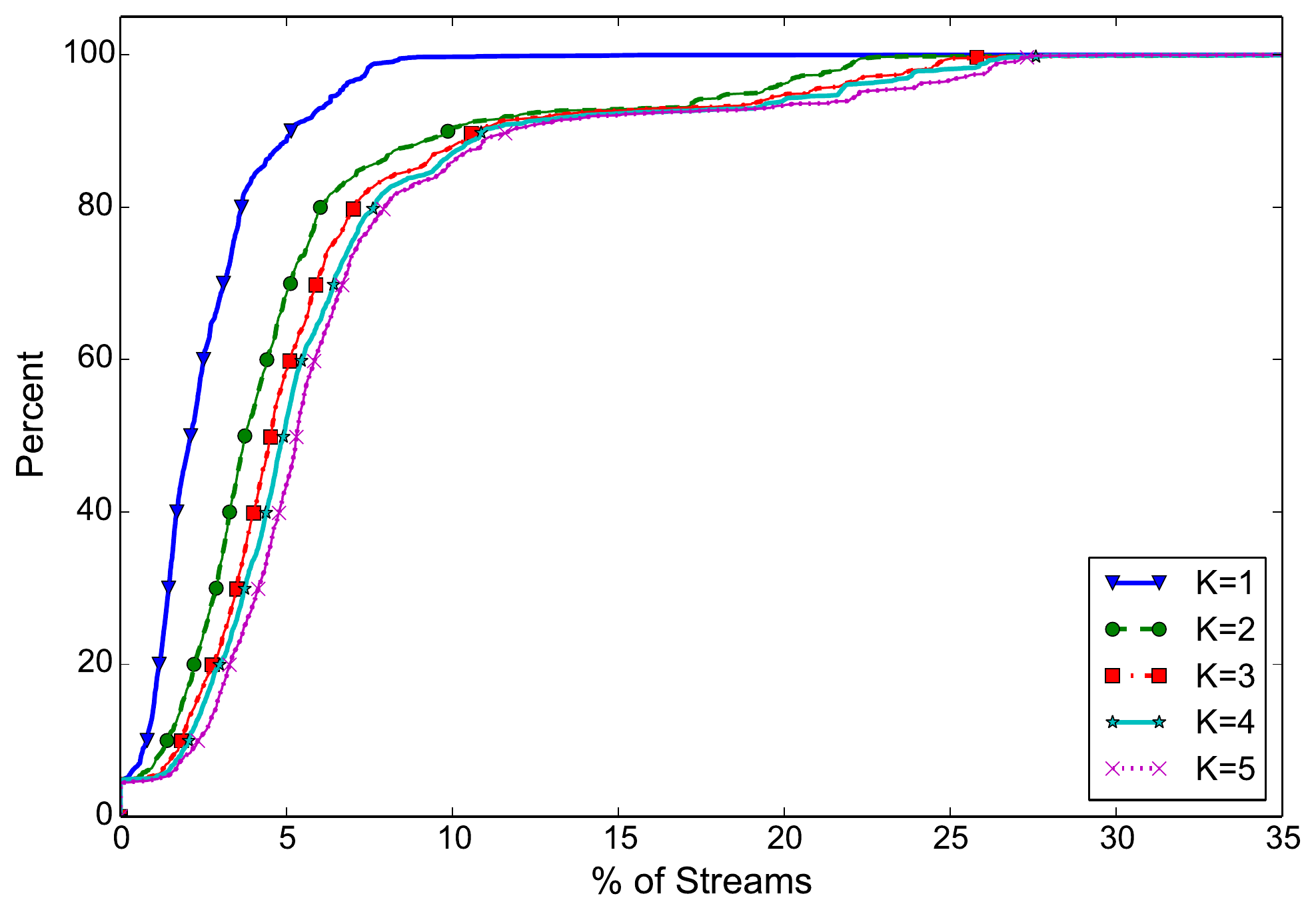}
\caption{AS Compromise Agreement}
\label{fig:asp}
			\end{minipage}
\hfill
			\begin{minipage}[b]{0.32\textwidth}
\includegraphics[width=\columnwidth]{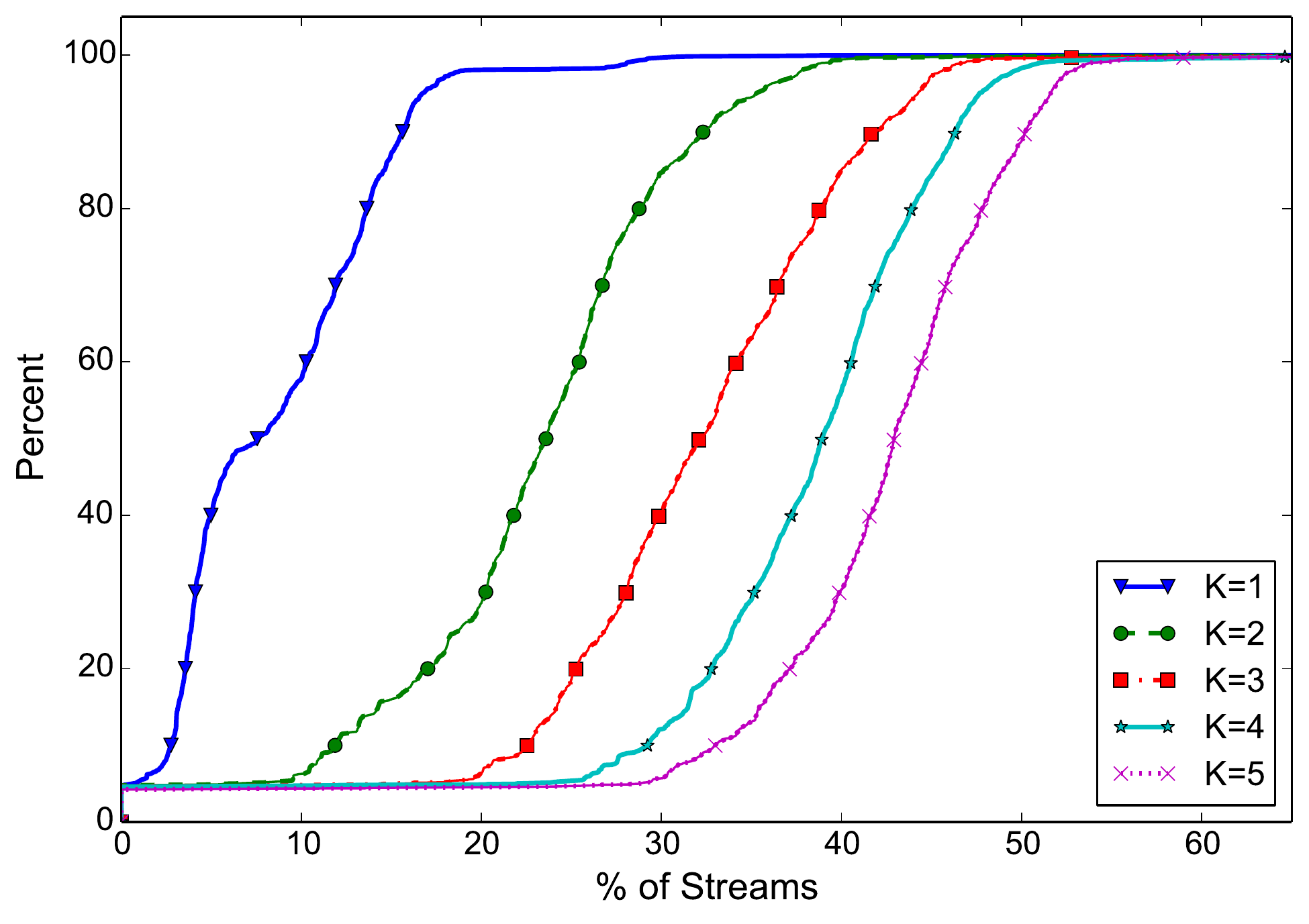}
\caption{AS Compromise False Positives}
\label{fig:asfp}
			\end{minipage}
\hfill
			\begin{minipage}[b]{0.32\textwidth}
\includegraphics[width=\columnwidth]{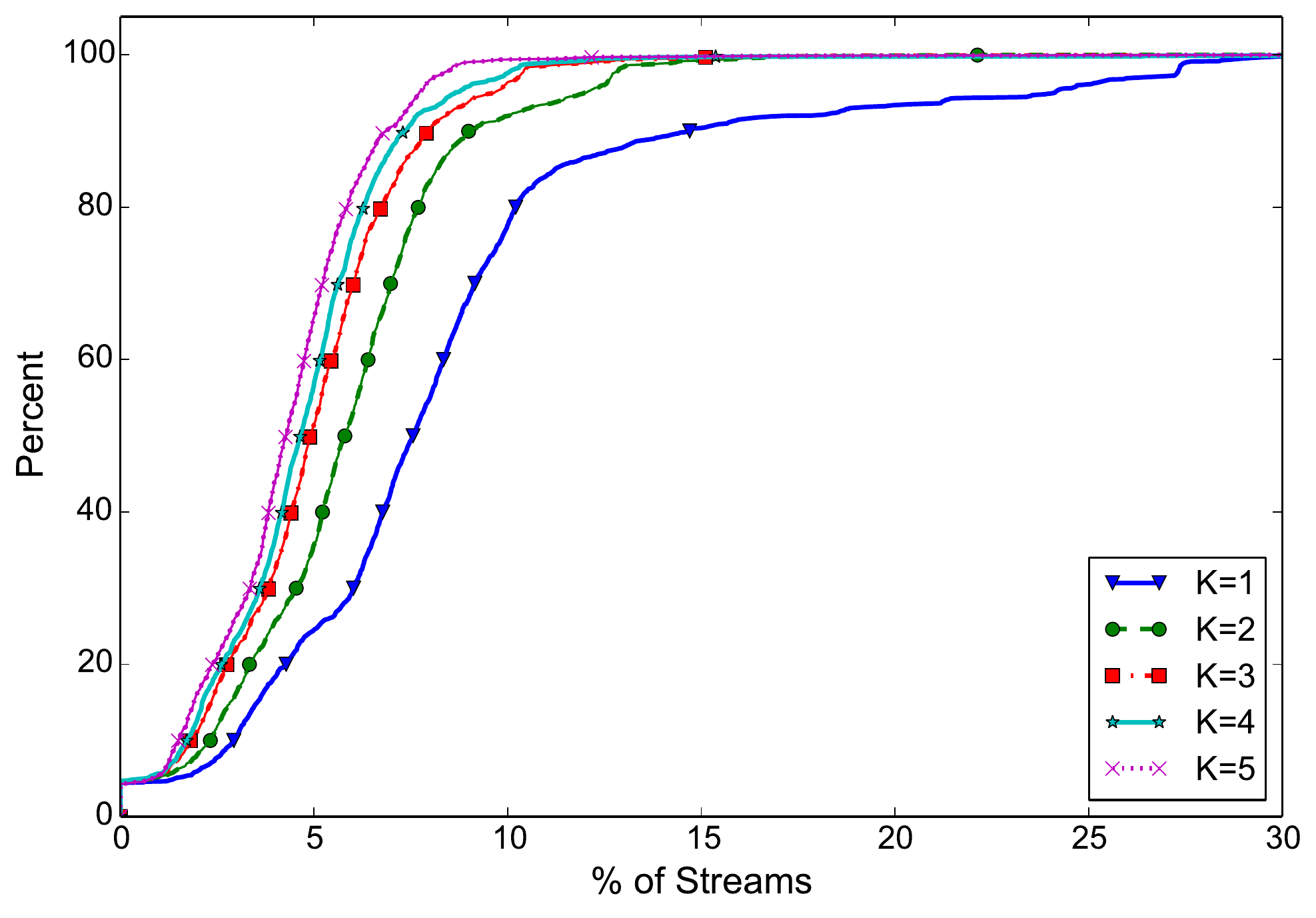}
\caption{AS Compromise False Negatives}
\label{fig:asfn}
			\end{minipage}
		\end{minipage}
		\begin{minipage}[t]{\textwidth}
			\begin{minipage}[b]{0.32\textwidth}
\includegraphics[width=\columnwidth]{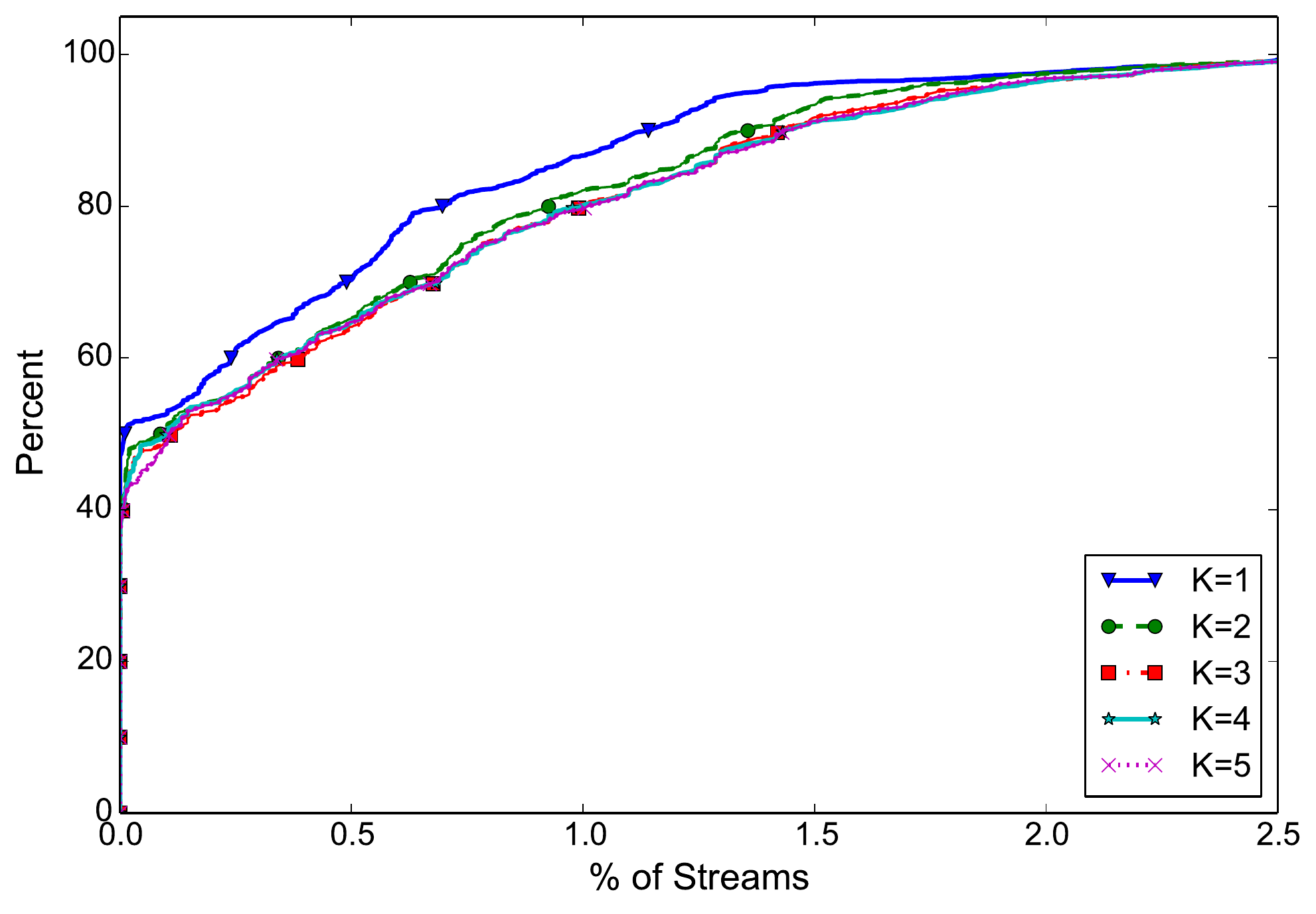}
\caption{IX Compromise Agreement}
\label{fig:ixp}
			\end{minipage}
\hfill
			\begin{minipage}[b]{0.32\textwidth}
\includegraphics[width=\columnwidth]{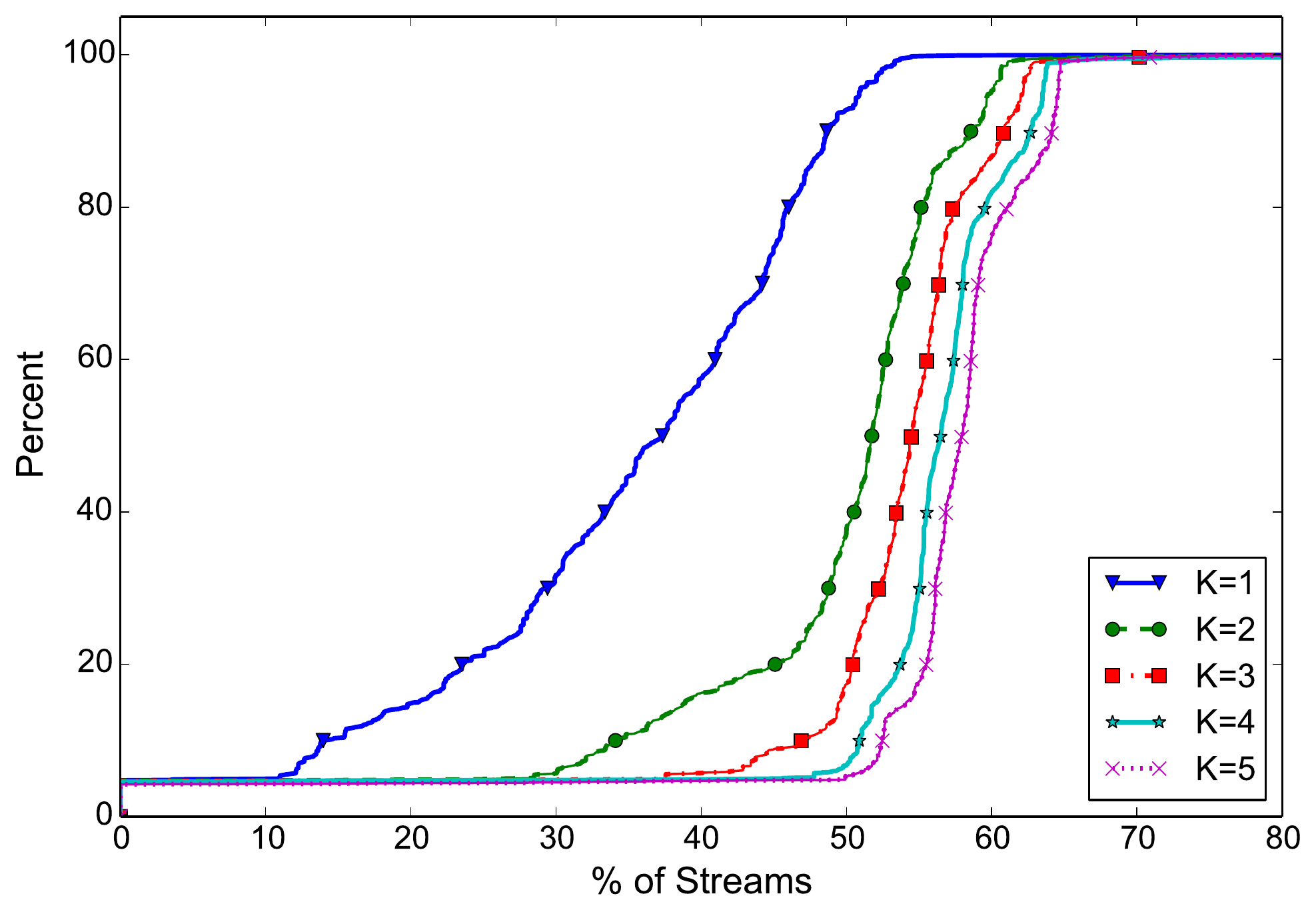}
\caption{IX Compromise False Positives}
\label{fig:ixfp}
			\end{minipage}
\hfill
			\begin{minipage}[b]{0.32\textwidth}
\includegraphics[width=\columnwidth]{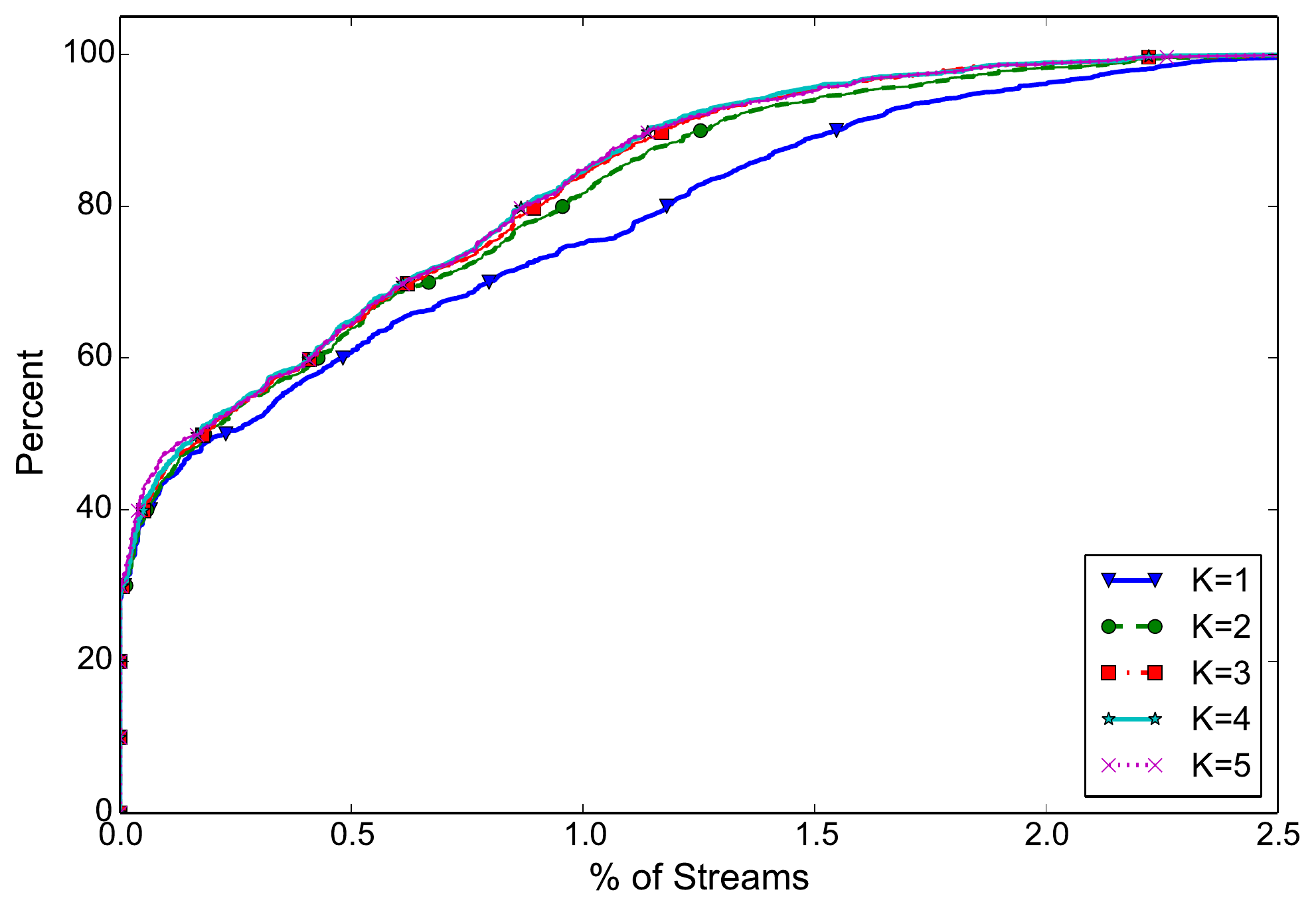}
\caption{IX Compromise False Negatives}
\label{fig:ixfn}
			\end{minipage}
		\end{minipage}
	\center{AS/IX Compromise CDFs Per Prefixes with both TraceRoute and Inferred Data}
	\end{minipage}
\end{figure*}

\subsubsection{Measured and Inferred Adversaries}

We now compare the inferred AS and IX adversaries to the AS and IX adversaries actually present in the traceroute measurements for all of our simulated Tor circuits. To make the comparison fair, we only consider the traceroutes and inferred paths going from the Tor guard to the client and from the Tor exit to the destination. As seen in the last section, the inferred paths using this Tor direction contain similar compromise rates to the paths in the forward and reverse directions. We thus consider the subset of paths for which we have both AS inferences and measured traceroutes in the Tor direction giving us a set of 141 million streams from 1000 unique client prefixes.

Figure \ref{fig:ascomprates} shows the CDF of streams compromised in the traceroute measurements compared to the inferred for various k top paths. Interestingly, the AS compromise rates for the top path is similar to the actual compromise rates seen in the measurements. Considering the top 2 paths more than doubles the inferred compromise rate with lower increases with increasing k topping out at a little under a 50\% compromise rate for half the paths. Figure \ref{fig:ixcomprates} shows the CDF of streams compromised with measured versus inferred IX adversaries. The actual percentage of paths with an IX adversary identified by prefix is much smaller than the inferred value with only .8\% of streams seeing an IX adversary on both the client to guard and exit to destination simultaneously. The inferred paths greatly over exaggerate the threat with the top path giving an average of 40\% compromise rate and the top 5 five paths giving an average of nearly 60\% compromise rate. Thus, the method of inferring IX adversaries greatly over predicts the number of actual IXes seen when measuring paths using traceroute. 

We now consider the differences between adversaries seen using the inference methods versus the adversaries seen in the traceroutes. We consider adversaries seen in the inferred set but not in the measured set as false positives and adversaries seen in the measurements but not the inferred set false negatives. While the traceroute measurement can contain errors and does not constitute perfect ground truth, we consider it more reliable than the inferred methods. In the following analysis all percentages are the percentage of paths compared to the set of all 141 million paths for which we have both inferred and measured data. 

Figures \ref{fig:asp} thru \ref{fig:ixfn} show the CDFs for the percentage of streams compromised per prefix for both ASes and IXes that both methods agree, the inference indicates an adversary while the measurement does not (false positives) and the measurement indicates an adversary while the inference does not (false negative). We see that while the percentage of overall AS compromises for the top path was similar in the last section, they do not agree on which AS is causing the compromise. In our measurements, we find roughly 10.9 \% of streams could contain a potential AS adversary. Unfortunately, the measured and inferred AS only agree for an average of 2.6\% AS compromises when considering the top path. Increasing to the top two paths improves this by a factor of 2 to 5.1\% average agreement with higher k values giving diminishing returns after that. Unfortunately, increasing k from 1 to 2 significantly increases the average number of false positives from 8.5\% to 22.6\% with a more linear increase with k up to 41.1\% with k = 5. For false negatives, the greatest drop once again occurs when going from the top paths to the top two paths from 8.4\% to 5.8\% with diminishing returns with increasing k and a minimum of 4.3\% average with k = 5. Overall, the AS inference with the top 2 paths catch a little less than half the measured AS adversaries catching 5.1\% and missing 5.8\% of the actually measured 10.9\% of measured AS adversaries Unfortunately, it still pre-emptively would eliminate 22.6\% of paths which had no measured adversary.

The agreement with the IX adversaries is even lower. Both methods agree on only .36\% of paths having an IX adversary considering the top path increasing to .44\% for the top two paths up to .47\% for the top 5 paths. The false positive rate is unacceptably high with 34.5\% for the top path and 48.1\% for the top 2 paths up to 55.7\% for the top 5 paths. The false negative rate is .54\% for the top path lowering to .45\% for the top two paths down to .42\% for the top five paths. Thus eliminating paths based on the inference with the top path would catch 40.0\% of the observed IX adversaries (.9\% of total paths) while eliminating 34.5\% of paths unnecessarily. Using the top 2 paths would catch 48.9\% of observed IX adversaries while eliminating 48.1\% of paths unnecessarily. Thus, roughly half of all potential paths would be eliminated to catch the .9\% of total paths with an observed IX adversary. This motivates the need for better methods of inferring IX adversaries in order to effectively mitigate the threat to the Tor system.

\begin{table*}[t]
\centering
\subfloat[Undetected compromise among streams that successfully connected]{
    \begin{tabular}{|l|r|r|}
    \hline & & \\ [-1.5ex]
     & \begin{tabular}{c}Top 1\\Path\end{tabular} & \begin{tabular}{c}Top 3\\Paths\end{tabular} \\
    \hline & & \\ [-1.5ex]
    \begin{tabular}{l}Mean fraction of streams\\that have traceroutes\end{tabular} & 0.0037 & 0.0026\\
    \hline
    \begin{tabular}{l}Mean fraction of streams\\with traceroutes that\\are w/o independence\end{tabular} & 0.0043 & 0.0014\\
    \hline
    \begin{tabular}{l}Min prob of at least one\\stream w/o independence\end{tabular} & 0.018 & 0\\
    \hline
    \begin{tabular}{l}Mean prob of at least one\\stream w/o independence\end{tabular} & 0.11 & 0.053\\
    \hline
    \begin{tabular}{l}Max prob of at least one\\stream w/o independence\end{tabular} & 0.22 & 0.18\\
    \hline
    \end{tabular}
    \label{table:pathindsec}
}
\qquad\qquad\qquad
\subfloat[Unnecessary failure among streams without any independent path]{
    \begin{tabular}{|l|r|r|}
    \hline & & \\ [-1.5ex]
     & \begin{tabular}{c}Top 1\\Path\end{tabular} & \begin{tabular}{c}Top 3\\Paths\end{tabular} \\
    \hline & & \\ [-1.5ex]
    \begin{tabular}{l}Mean fraction of all\\streams that fail due to\\independence constraint\end{tabular} & 0.051 & 0.060\\
    \hline
    \begin{tabular}{l}Mean fraction of streams\\that have traceroutes\end{tabular} & 0.19 & 0.19\\
    \hline
    \begin{tabular}{l}Mean fraction of streams\\w/ traceroutes that have\\an independent path\end{tabular} & 0.96 & 0.95\\
    \hline
    \begin{tabular}{l}Min prob of at least\\one stream failure\end{tabular} & 1 & 1\\
    \hline
    \end{tabular}
    \label{table:pathindfailures}
}

\caption{Path-independent Tor traceroute analysis over 189 top client ASes}
\label{table:pathind}

\end{table*}

\pagebreak
\subsection{Path-independent Tor}
In order to avoid deanonymization by an AS or IX, Tor clients could attempt to choose Tor relays
such that the forward and reverse paths between the client and guard are independent of the forward
and reverse paths between the exit and destination, in terms of the ASes and IXes that appear.
However, it is non-trivial to design a system that allows the client to do so, because he must
preserve his anonymity while making this decision, and Tor should be usable even by users with
little bandwidth and low-powered devices.

As we discuss in Section~\ref{sec:related}, Edman and Syverson~\cite{ccs-EdmanS09} presented the
first detailed proposal for solving this
problem with a system that provides enough data for clients to build an AS Internet map on which
to run AS-path inference. They propose a slightly \emph{less} accurate algorithm than Qiu
and Gao's for efficiency. Juen added IX inference to this idea~\cite{anonymity-ipx-thesis}. None of
this previous work explains how AS/IX-independent circuits should be created \emph{over time},
and thus does not consider how path independence interacts with Tor guards or circuit reuse. Tor
guards in particular are a key Tor feature that defends against malicious
observation and deanonymization~\cite{cogs,usersrouted-ccs13}. Thus the prior work does not give a
clear idea of how well AS/IX-independent path selection would work even if path-inference
techniques were very accurate.

The inaccuracy of path-inference techniques is likely to negatively
impact AS/IX-independent path selection in at least two ways:
(\emph{i}) missing an AS or IX on a path could cause the user to create a path vulnerable to
deanonymization, and (\emph{ii}) incorrectly believing that an AS or IX exists on a path could
leave the user
with few or no ways to connect to the destination. These problems are placed in tension by the
inference methodology because false negatives make (\emph{i}) worse and false positives make
(\emph{ii}) worse. For example, as the number $k$ of top paths used in inference increases,
false negatives should go down but false positives should go up. Moreover, the inference needs to
have few false negatives on all paths \emph{collectively}, or a user will face an increasing risk of
deanonymization as he visits new destinations and is forced by network churn to use
different relays. Similarly, an increasing number of false positives over time could force the user
to choose between not connecting to certain destinations and exposing himself to more and more
potentially-malicious guards.

We investigate the suitability of path inference as a basis for AS/IX-independent path selection
using path simulation and our traceroute data, similar to how they were used in
Section~\ref{subsec:vanilla} to explore vanilla Tor
security. As a byproduct of this research, we also expose a security-performance tradeoff inherent
in the path-independent approach and reveal some opportunities to fill in and improve past
proposals.

\subsubsection{Methodology}
In order to evaluate AS/IX-independent path selection via simulation, we must fill in the details
of the algorithm sketched out by prior work. We adapt the existing Tor path-selection algorithm for
this purpose. We require clients to have at least 3 guards in their guard list and to have at least
2 guards active with AS/IX path information with the client when creating a new circuit. Upon
receiving a stream request, existing circuits are examined for suitability, including path
independence. If
none is suitable, then circuit-creation is initiated by choosing an existing guard, then an exit,
and then a middle. If a path-independent exit cannot be found for a given guard, the other guards
are considered, and so on. If no exit is found for any current guard, then the circuit creation
fails. We note that to enable a direct comparison with our traceroute data, our simulator only
compares the inferred AS/IX path from the guard to the client and
from the exit to the destination when determining path independence (i.e. only reverse entry and
forward exit paths are used).

We generally follow the same experimental methodology as that followed in
Section~\ref{subsec:vanilla}. We will not be estimating full distributions, and thus we use only
500 samples per client AS, but we run experiments with the 189 of the top 200 client ASes that were
in our AS-level routing map.

In our experimental analysis, we are able to use traceroute data to identify false
negatives and false positives. 50 client IPs are chosen randomly from the set of the initial IPs in
each prefix advertised by the client's AS (according to the RouteViews prefix-to-AS file that
appears most recently before that stream occurred). To identify false negatives, we test streams 
that were successfully assigned to a circuit by looking for a traceroute from the guard's routing
prefix to the client's and from the exit's prefix to the destination's. When both traceroutes are
found, we look for ASes or IXes that appear in common. To identify false positives, we test streams
that failed to connect by looking for a traceroute from any of the active guards at that time to the
client and from any potential exit to the destination. If such a pair exists, we look for the lack
of any AS or IX in common.

\subsubsection{Results}

Table~\ref{table:pathindsec} provides estimates for the effects of path inference errors on the 
security of path-independent Tor. The min, mean, and max values are taken over 188 top client ASes
(we further excluded one that didn't advertise any prefixes during the simulation week). Our
traceroute data provided path information (i.e. matched
both guard-client and exit-destination host-prefix pairs in the direction out from Tor) for
0.26--0.38\% of the simulations' streams (depending on whether the top 1 or the top 3 inferred paths
were used to determine independence). Of these, between 0.14\% and 0.43\% were revealed tofigures/
violate path independence. While this may seem acceptably low, even one Tor deanonymization is
potentially serious, and over the course of the simulated week, a client had on average between
a 5.3\% and a 11\% probability of experiencing at least one path-independence violation. In the most
unlucky client ASes, path independence was violated with a probability as high as 18--21\%!

Table~\ref{table:pathindfailures} shows that this insecurity cannot simply be handled by increasing
the number of top possible paths from which the inferred ASes and IXes are taken. It reveals that
by increasing the number of top paths used in inference from 1 to 3, the fraction of streams for
which no path-independent Tor circuit could be created increased from 5.1\% to 6\%. For these
streams, no AS/IX path-independent exit could be found using any of the client's guards.
Note that a stream failure of any kind \emph{never} occurred in simulation with Tor's default path
selection, because Tor doesn't require path independence, and many exits are available for each 
stream in the user trace. Such failures are particularly bad because the stream will not succeed
until the Tor relay population changes sufficiently, a process which could take days or weeks. Thus
even a 5.1--6\% failure rate has a severely deleterious effect on Tor's suitability for general
Internet use. Moreover, we can see that every simulated client experienced at least one stream
failure (i.e. the estimated failure probability is 1.0 for all client ASes).

However, our traceroute measurements offer the hopeful news for this problem that most of these
stream failures may have been unnecessary. We were able to match a traceroute guard-to-client and
exit-to-destination for 19\% of failed streams. Our coverage of failed streams is so much
higher than for connected streams
because we look for a traceroute from any active guard of the client at the time and from any exit
that could be chosen with that guard and for that destination (ignoring only the path-independence
constraint). Among streams for which we were able to match at least one pair of traceroutes,
95--96\% had a guard and exit that the traceroutes show would have been AS/IX-independent. In fact,
this high false-positive rate is not just a result of having many exit paths about which to be
be incorrect --- an average of about 80\% of all guard and exit pairs with matched outgoing 
traceroutes were observed to be path-independent for both experiments.


\subsubsection{Discussion}
Our evaluation of AS/IX-independent path selection is not intended to make any definitive claims
about its usefulness. Instead, we attempt to make reasonable choices about the algorithm details
in order to get some idea of how well it might work overall and especially in conjunction with
path-inference techniques. Indeed, there are many plausible improvements to the algorithm we have
evaluated, such as choosing guards with different network locations to minimize the chance of stream
failure, or perhaps allowing streams to use potentially unsafe circuits but limiting the number
of potential observing ASes and IXes. Designing network-aware path-selection algorithms for Tor
remains an open challenge with unsolved vulnerabilities such as adversarial relay
placement~\cite{lastor} and path fingerprinting~\cite{danezis-pets2008,trust-ccs11}.

%% file: related.tex
\section{Related Work}  \label{sec:related}

The threat to the Tor network for ASes to correlate traffic was first investigated by Feamster and Dingledine \cite{feamster:wpes2004}. Using a simplified AS model with shortest paths they determined roughly 10-30\% of circuits could be vulnerable to an AS adversary. Edmond and Syverson furthered the understanding of AS adversaries against the Tor network \cite{ccs-EdmanS09}. Using Qiu and Gao's AS path prediction model and an updated model for the Tor network, they determined each circuit had an 11-18\% chance that some AS adversary could compromise the circuit. They also presented a technique to choose paths without AS adversaries by using "Snapshots'' of the AS topology. Akhoondi et al. presented LastTor, an optimization to Tor path selection to minimize latency by considering geographic location \cite{lastor}. They propose using the set of K top most likely AS paths to eliminate AS adversaries. They do not report overall chances for any given AS to compromise a circuit. Recently, Wacek et al. studied Tor's path selection algorithm \cite{wacek2013empirical}. They find that using the iPlane's Nano AS map, Tor paths have a 27.39\% chance to be vulnerable to an AS adversary. 

The danger of IX adversaries was first demonstrated by Murdoch and Zielinski who demonstrated that an IX could use a Bayesian approach to sample traffic and correlate Tor flows across ASes peering at the IX \cite{murdoch-pet2007}. Juen further investigated the threat of AS and IX adversaries using Qiu and Gao's AS model and the top K paths estimating the chance of any AS being able to compromise the circuit ranging from 10\% to 42\% \cite{anonymity-ipx-thesis}. He reports the chance of an IX compromise to be between 1 \% and 20 \%. Johnson et al. investigate the amount of time required for an AS, IX, or IX organization to compromise a circuit using Torps to simulate realistic Tor traffic \cite{usersrouted-ccs13}. They only consider the top 3 AS and IX adversaries as seen in their inferred data and report the overall chance of an AS compromise to be 1.6 \% for their top 3 ASes.

\begin{table}
	\centering
	\begin{tabular}{|l|ccc|}
	\hline & & & \\ [-1.5ex]
	Method & Forward & Reverse & Both \\
	\hline & & & \\ [-1.5ex]
	Feamster and Dingledine & 17.7\% & 16.1\% & NA \\
	Edmond and Syverson & 10.9\% & 11.1\% & 17.8\% \\
	Wacek et al. & NA & NA & 27.39 \% \\
	Juen & 7.1 \% & 7.2 \% & 11.2 \% \\
	Current Work & 11.6\% & 12.1\% & 21.6\% \\ 
	\hline
	\end{tabular}
\caption{Inferred AS Compromise Comparison (Top Path)}
\label{table:infASComp}
\end{table}

\begin{table*}
	\centering
	\begin{tabular}{|l|cc|ccc|}
	\hline & & & & & \\ [-1.5ex]
	AS & Our Rank & Johnson et al. Rank & Johnson et al. Comp \% & Comp \% &  TR Comp \%  \\
	\hline & & & & & \\ [-1.5ex]
	AS6939 HURRICANE Electric & 1 & 3 & .6\% & .4\% & 0.0\% \\
	AS3356 Level 3 Communications & 2 & 1 & .4\% & .5\% & .13\% \\
	AS1299 TeliaNet Global & 3 & 2 & .4\% & .5\% & .5 \%\\
	\hline & & & & & \\ [-1.5ex]
	IX & Our Rank & Johnson et al. Rank & Johnson et al. Comp \% & Comp \% & TR Comp \% \\
	\hline & & & & & \\ [-1.5ex]
	LINX Juniper & 1 & NA & NA & .4\% & .05\%\\
	DE-CIX Frankfurt & 2 & 1 & .1 \% & .4\% & .05\%\\
	Equinix Ashburn & 3 & NA & NA & .4\% & 0.0\%\\
	\hline
\end{tabular}
\caption{Stream Compromise Rates for the Top 3 AS and IX Adversaries for our Work compared to Johnson et al.}
\label{table:topasix}
\end{table*}

We now compare our results with the compromise rates of Tor streams against previous work. We calculate the percentage of Tor streams which contain an AS on the client to guard and exit to destination paths in the forward, reverse and forward and reverse directions for each of our 18 billion calculated streams. We then compare the results of our directional AS path inferences directly to the results from previous work and confirm that our AS path inferences give similar results for the top AS path as shown in Table \ref{table:infASComp}. We find our results most closely correlate with the work of Edmond and Syverson with Juen's results being lower than the average and Wacek et al. being much higher. We find this unsurprising since we also use the AS inference algorithm from Qiu and Gao. We surmise that the AS inference from iPlanes produce higher compromise estimates as seen in Feamster and Dingledine and Wacek et al. Juen also uses a modified AS mapping algorithm which may produce lower compromise rates.

Johnson et al. investigated the time expected before a user would most likely use a stream compromised by an AS or IX adversary. Since we only have inferred and traceroute data for .8\% of streams, it is not possible to directly compare the time to compromise for our clients. Instead, we investigate the ability of the top 3 AS and IX adversaries to compromise a Tor stream. Once again, we only consider streams which we have both inferred and traceroute data. The ASes and IXes with the highest probability to compromise a Tor stream are shown in Table \ref{table:topasix}. Interestingly, we observe the same set of three top AS adversaries but in a slightly different order. We also see the top IX adversary as number 2 for compromise rates. We see a similar .5\% rate of AS streams compromised by our top AS adversaries. We see a higher rate of overall streams compromised by our top IX adversary at .4\% compared to roughly .1\% in Johnson's work. The overall compromise rates of all streams using the traceroute measurements is much more interesting. The traceroute measurements never see AS6939, the top compromising AS in the inferences. On the other hand, the traceroute measurements indicate AS1299 can compromise the same percentage of streams as indicated in the inference. We see drops for both top IX compromise rates for the measurements versus the inferences. Once again, Equinix Ashburn is not observed in the traceroute measurements.

Overall, we expect a drop in the actual ability of an IX point to compromise a Tor stream primarily due to the extremely high false inference of IX points. The difference between AS compromise rates is more interesting. This shows that the path inference was always wrong in identifying AS6939 when compared to traceroutes; however, AS 1299 was seen in the traceroutes at similar rates as predicted. Thus, the inference accuracy appears to vary greatly depending on which AS is being considered as an adversary.

%% file: conc.tex
\section{Limitations and Future Work}\label{sec:future}
This data in this study is limited in several ways.
While our volunteer measuring relay covered roughly 25\% of Tor selection probability, it still only
contains 28 hosts. In addition, all path inferences were done
on paths from Tor relays, leaving us without symmetric path information. Furthermore, we collected
most of our data in the span of weeks, and so missed alternative routing
paths and routing instabilities. We also lack ground truth because of measurement weaknesses
such as missing or incorrect traceroute hops, missing or stale IP prefix announcements from the
public route collectors, and incomplete or
incorrect IXP prefix data. We look forward to the opportunity to expand network measurement in cooperation with Tor and using third-party vantage points such as Looking Glass servers.
We also hope to make use of advances in measurement tools to advance this line of inquiry.

A larger remaining challenge is to move past the current focus on AS-level techniques and
adversaries. As Jaggard et al.
describe~\cite{trustrep-hotpets14}, an adversary may find it easier to control a group of IP routers
running a certain version of software than to observe those in the same
AS or IXP. Evaluating the threat of more complicated and realistic network adversaries will require
both better adversary modeling and more detailed route inference techniques.

\section{Conclusions}\label{sec:conclusion}       

We have presented a measurement study to evaluate the suitability of Internet AS and IXP
path-prediction
algorithms to assess and mitigate the threats from network-level adversaries to the Tor network.
Using traceroute data from the volunteer operators of 28 Tor relays, we show that current techniques
for inferring AS-level Internet paths and the IXes between them significantly overestimate number of ASes and IXes traversed by Tor traffic.

To evaluate what this means about the current and future security of Tor, we perform
Monte Carlo simulations of Tor's current path-selection algorithm and the AS/IXP-independent
path-selection algorithm proposed in the literature.
When we examine the results, we see evidence that Tor is likely less vulnerable to an AS or IXP
adversary than has been previously found. A direct comparison
with a prior evaluation shows that it is likely to have overstated the risk of a single AS many
times over and that of a single IXP by an order of magnitude.

We also find that the AS/IXP-independent path-selection algorithm
may still leave a significant chance for users to be deanonymized over time
due to the errors in path prediction --- we estimate a 5--11\% risk in just one week when the
claimed chance is 0.
Moreover, we find that this algorithm appears to force a tradeoff between connection failures and
exposing users to potentially-malicious relays,
even though in nearly all cases the failures could be avoided with better measurement.

Our results suggest the importance of accurate measurement both for understanding Tor security
and for improving it.